\documentclass[12pt,letterpaper]{article}
\usepackage{epsfig,setspace,amsmath,epsf,amssymb,bm,theorem,cite, graphicx, epstopdf, algorithm, algpseudocode,float,color}
\usepackage{multirow}
\usepackage[table,xcdraw]{xcolor}

\newtheorem{theorem}{Theorem}
\newtheorem{corollary}{Corollary}

\newtheorem{lemma}{Lemma}

\newenvironment{Proof}[1]{\medskip\par\noindent{\bf Proof:\,}\,#1}{{\mbox{\,$\blacksquare$}\par}}

\allowdisplaybreaks

\begin{document}
	
\title{Fundamental Limits of Cache-Aided Private Information Retrieval with Unknown and Uncoded Prefetching\thanks{This work was supported by NSF Grants CNS 13-14733, CCF 14-22111, and CNS 15-26608.}}
	
\author{Yi-Peng Wei \qquad Karim Banawan \qquad Sennur Ulukus\\
	\normalsize Department of Electrical and Computer Engineering\\
	\normalsize University of Maryland, College Park, MD 20742 \\
	\normalsize {\it ypwei@umd.edu  \qquad \it kbanawan@umd.edu} \qquad {\it ulukus@umd.edu}}
	
\maketitle
	
\vspace*{-0.8cm}

\begin{abstract}
We consider the problem of private information retrieval (PIR) from $N$ non-colluding and replicated databases when the user is equipped with a cache that holds an uncoded fraction $r$ from each of the $K$ stored messages in the databases. We assume that the databases are unaware of the cache content. We investigate $D^*(r)$ the optimal download cost normalized with the message size as a function of $K$, $N$, $r$. For a fixed $K$, $N$, we develop an inner bound (converse bound) for the $D^*(r)$ curve. The inner bound is a piece-wise linear function in $r$ that consists of $K$ line segments. For the achievability, we develop explicit schemes that exploit the cached bits as side information to achieve $K-1$ non-degenerate corner points. These corner points differ in the number of cached bits that are used to generate one side information equation. We obtain an outer bound (achievability) for any caching ratio by memory-sharing between these corner points. Thus, the outer bound is also a piece-wise linear function in $r$ that consists of $K$ line segments. The inner and the outer bounds match in general for the cases of very low caching ratio ($r \leq \frac{1}{1+N+N^2+\cdots+N^{K-1}}$) and very high caching ratio ($r \geq \frac{K-2}{(N+1)K+N^2-2N-2}$). As a corollary, we fully characterize the optimal download cost caching ratio tradeoff for $K=3$. For general $K$, $N$, and $r$, we show that the largest gap between the achievability and the converse bounds is $\frac{1}{6}$. Our results show that the download cost can be reduced beyond memory-sharing if the databases are unaware of the cached content.
\end{abstract}

\section{Introduction}
%PIR problem, trivial scheme, retrieval rate
The problem of private information retrieval (PIR) was introduced by Chor et al. \cite{ChorPIR} as a canonical problem to investigate the privacy of the contents downloaded from public databases. The PIR problem has become a major research area within the computer science literature subsequently, see e.g., \cite{PIRsurvey2004,cachin1999computationally,ostrovsky2007survey,yekhanin2010private}. In the classical form of the problem \cite{ChorPIR}, a user requests to download a message (or a file) from $K$ messages from $N$ non-communicating databases such that no database can distinguish individually which message has been retrieved. The user performs this task by preparing $N$ queries, one for each database, such that the queries do not reveal the user's interest in the desired message. Each database responds truthfully to the received query by an answer string. The user reconstructs the desired message from the collected answer strings. A feasible PIR scheme is to download all of the $K$ messages from a database. However, this trivial PIR scheme is quite inefficient from the retrieval rate perspective, which is defined as the number of desired bits per bit of downloaded data. Consequently, the aim of the PIR problem is to retrieve the desired message correctly by downloading as few bits as possible from the $N$ databases under the privacy constraint.      

%Computer science formulation, information theoretic formulation and literature
Recently, the PIR problem is revisited by information theorists \cite{RamchandranPIR,RazanPIR,YamamotoPIR,JafarPIR}. In the information-theoretic re-formulation of the problem, the length of the message $L$ is assumed to be arbitrary large to conform with the traditional Shannon-theoretic arguments, and the upload cost is neglected as it does not scale with the message length. This formulation provides an absolute privacy guarantee by ensuring statistical independence between the queries and the identity of the desired message. In the influential paper by Sun and Jafar \cite{JafarPIR}, the notion of PIR capacity is introduced, which is the supremum of PIR rates over all achievable retrieval schemes. Reference \cite{JafarPIR} characterizes the capacity of classical PIR. In \cite{JafarPIR}, a greedy iterative algorithm is proposed for the achievability scheme and an induction based converse is provided to obtain an exact result. The achievable scheme is based on an interesting correspondence between PIR and blind interference alignment \cite{BIA} as observed earlier in \cite{JafarPIRBlind}. Sun and Jafar show that in order to privately retrieve a message, the optimal total downloaded bits normalized with the message size is $\frac{D}{L}=1+\frac{1}{N}+\dots+\frac{1}{N^{K-1}}$. Consequently, the PIR capacity is the reciprocal of this optimal normalized download cost, i.e., $C=(1+\frac{1}{N}+\dots+\frac{1}{N^{K-1}})^{-1}$. 

Following the work of \cite{JafarPIR}, the fundamental limits of many interesting variants of the classical PIR problem have been considered, such as: PIR with $T$ colluding databases (TPIR) \cite{JafarColluding,arbitraryCollusion}, where any $T$ of $N$ databases might collude; robust PIR (RPIR) \cite{JafarColluding}, where some databases may fail to respond; symmetric PIR (SPIR) \cite{symmetricPIR}, which adds the constraint that the user should only learn the desired message; MDS-coded PIR (CPIR) \cite{KarimCoded}, where the contents of the databases are not replicated, but coded via an MDS code; multi-message PIR (MPIR) \cite{MPIRjournal}, where the user wishes to jointly retrieve $P$ messages; PIR from Byzantine databases (BPIR), where $B$ databases are outdated or worse adversarial \cite{BPIRjournal}; PIR under message size constraint $L$ (LPIR) \cite{arbmsgPIR}; multi-round PIR, where the queries are permitted to be a function of the answer strings collected in previous rounds \cite{MultiroundPIR}; MDS-coded symmetric PIR \cite{codedsymmetric}; MDS-coded PIR with colluding databases \cite{codedcolluded, codedcolludedJafar, codedcolludingZhang}, and its multi-message \cite{MPIRcodedcolludingZhang}, and symmetric \cite{wang2017linear} versions. 
% Caching literature % Ravi's Paper

Recently, reference \cite{tandon2017capacity} has considered cache-aided PIR, where the user has local cache memory of size $rKL$ bits and it can store any function of the $K$ messages subject to this memory size constraint\footnote{Caching is an important technique to reduce the peak-time traffic in networks by pre-storing (prefetching) content to end-user's local memory \cite{maddah2014fundamental,maddah2015decentralized}, and has been an active recent research field on its own right.}$^{,}$\footnote{In another related line of work, reference \cite{privateBC_karmoose} investigates the privacy risks when the clients of an index coding based broadcast system possess a subset of the messages as side information and use them to retrieve the desired message privately against an external eavesdropper.}. With the assumption that the cache content is known by all the $N$ databases, reference \cite{tandon2017capacity} characterizes the optimal download cost. The achievability scheme is based on memory-sharing and the converse bound is obtained with the aid of Han's inequality. To privately retrieve a message, the optimal total downloaded bits normalized with the message size is $\frac{D(r)}{L}=(1-r)(1+\frac{1}{N}+\dots+\frac{1}{N^{K-1}})$. The result is quite pessimistic as it implies that the cached bits cannot be used as side information within the retrieval scheme and the user must download the uncached portion of the file (the remaining $L(1-r)$ bits) using the original PIR scheme in \cite{JafarPIR}. The reason behind this result is that the databases are fully knowledgeable about the cached bits and can infer which message is desired if the user exploits these cached bits as side information in any form.
% Motivation to our problem, system model

The above discussion motivates us to investigate the other extreme where the databases are fully unaware of the cache content, i.e., when the prefetched bits are unknown to all of the $N$ databases (in contrast to having the cache content as public knowledge at all the $N$ databases as in \cite{tandon2017capacity}). In this case, the user can leverage the cached bits as side information without sacrificing the privacy constraint as the databases are unaware of the cached bits. This poses an interesting question: What is the optimal way to exploit the cached bits as side information in order to minimize the normalized download cost, and what is the corresponding gain beyond memory-sharing if any? The assumption of unknown prefetching can be interpreted in practice as either the prefetching phase is performed via an external database which does not participate in the retrieval (delivery) phase, or in the context of dynamic cache-aided PIR,
in which once the unknown cache is used, the user updates/refreshes its cached contents by some trusted mechanism which keeps the cached content essentially random from the perspective of each database as pointed out by \cite{tandon2017capacity}. In this paper, we further assume that the cache content is uncoded. As a main advantage, uncoded prefetching allows us to handle asynchronous demands without increasing the communication rates, by dividing files into smaller subfiles \cite{yu2016exact}. 

% This paper results
In this work, we consider PIR with unknown and uncoded prefetching, i.e., we assume that the cache content is unknown to all databases, and the cache supports only direct (uncoded) portions of all messages (smaller subfiles). We aim to characterize the
optimal tradeoff between the normalized download cost $\frac{D(r)}{L}$ and the caching ratio $r$. For the outer bound, we explicitly determine the achievable download rates for specific $K+1$ caching ratios. Download rates for any other caching ratio can be achieved by proper memory-sharing between the nearest two explicit points. This implies that the outer bound is a piece-wise linear curve which consists of $K$ line segments. For the inner bound, we extend the techniques of \cite{JafarPIR,tandon2017capacity} to obtain a piece-wise linear curve which also consists of $K$ line segments. We show that the inner and the outer bounds match exactly at three of the line segments for any number of messages $K$. This means that we characterize the optimal tradeoff for the very low ($r \leq \frac{1}{1+N+N^2+\cdots+N^{K-1}}$) and the very high ($r \geq \frac{K-2}{(N+1)K+N^2-2N-2}$) caching ratios. As a direct corollary, we fully characterize the optimal download cost caching ratio tradeoff for $K=3$ messages. For general $K$, $N$ and $r$, we show that for fixed $N$, the outer bound monotonically increases as $K$ increases. To characterize the worst-case gap between the inner and the outer bounds, we determine the asymptotic achievability bound as $K \rightarrow \infty$ for fixed $N$, $r$. We then show that the asymptotic gap monotonically decreases in $N$. Therefore, the worst-case gap happens at $N=2$ and $K \rightarrow \infty$. By maximizing this over $r$, we show that the largest gap between the achievability and the converse bounds is $\frac{1}{6}$. Our results show the benefits of the cached content when the databases are unaware of it over the scenario in \cite{tandon2017capacity} where the databases are fully aware of the cached content.

\section{System Model}
We consider a classic PIR problem with $K$ independent messages $W_1, \dots, W_K$. Each message is of $L$ bits long, 
\begin{align}
H(W_1)=\dots=H(W_K)=L, \qquad H(W_1, \dots, W_K)=H(W_1)+\dots+H(W_K). 
\end{align}
There are $N$ non-communicating databases, and each database stores all the $K$ messages, i.e., the messages are coded via $(N,1)$ repetition code \cite{KarimCoded}. The user (retriever) has a local cache memory whose content is denoted by a random variable $Z$. For each message $W_k$ of $L$ bits long, the user randomly and independently caches $Lr$ bits out of the $L$ bits to $Z$, where $0\leq r \leq 1$, and $r$ is called the \textit{caching ratio}. Therefore, 
\begin{align}
H(Z)=KLr.
\end{align}
Since the user caches a subset of the bits from each message, this is called \textit{uncoded prefetching}. We denote the indices of the cached bits by random variable $\mathbb{H}$. Here, different from \cite{tandon2017capacity}, we consider the case where none of the databases knows the prefetched cache content. 

After the uncoded prefetching phase, the user privately generates an index $\theta \in [K]$, where $[K]=\{1,\dots, K\}$, and wishes to retrieve message $W_\theta$ such that no database knows which message is retrieved. Note that during the prefetching phase, the desired message is unknown a priori. Note further that the cached bit indices $\mathbb{H}$ are independent of the message contents and the desired message index $\theta$. Therefore, for random variables $\theta$, $\mathbb{H}$, and $W_1,\dots,W_K$, we have
\begin{align} \label{independency}
H\left(\theta, \mathbb{H}, W_1,\dots,W_K  \right)= H\left( \theta \right) + H\left( \mathbb{H} \right) + H(W_1)+\dots+H(W_K). 
\end{align}

Suppose $\theta=k$. The user sends $N$ queries $Q_1^{[k]}, \dots, Q_N^{[k]}$ to the $N$ databases, where $Q_n^{[k]}$ is the query sent to the $n$th database for message $W_k$. The queries are generated according to $\mathbb{H}$, which is independent of the realizations of the $K$ messages. Therefore, we have 
\begin{align} \label{query_indep}
I(W_1, \dots, W_K; Q_1^{[k]}, \dots,  Q_N^{[k]}  ) =0. 
\end{align}
To ensure that individual databases do not know which message is retrieved, we need to satisfy the following privacy constraint, $\forall n \in [N]$, $\forall k \in [K]$, 
\begin{align} \label{privacy_constraint}
(Q_n^{[1]}, A_n^{[1]}, W_1, \dots, W_K) \sim (Q_n^{[k]}, A_n^{[k]}, W_1, \dots, W_K). 
\end{align}

Upon receiving the query $Q_n^{[k]}$, the $n$th database replies with an answering string $A_n^{[k]}$, which is a function of  $Q_n^{[k]}$ and all the $K$ messages. Therefore, $\forall k \in [K], \forall n \in [N]$, 
\begin{align} \label{answer_constraint}
H(A_n^{[k]}|Q_n^{[k]}, W_1, \dots, W_K)=0. 
\end{align}
After receiving the answering strings $A_1^{[k]}, \dots, A_N^{[k]}$ from all the $N$ databases, the user needs to decode the desired message $W_k$ reliably. By using Fano's inequality, we have the following reliability constraint 
\begin{align} \label{reliability_constraint}
H\left(W_k|Z, \mathbb{H}, Q_1^{[k]}, \dots, Q_N^{[k]}, A_1^{[k]}, \dots, A_N^{[k]} \right) = o(L), 
\end{align}
where $o(L)$ denotes a function such that $\frac{o(L)}{L} \rightarrow 0$ as $L \rightarrow \infty$. 

For a fixed $N$, $K$, and caching ratio $r$, a pair $\left(D(r),L\right)$ is achievable if there exists a PIR scheme for message of size $L$ bits long with unknown and uncoded prefetching satisfying the privacy constraint \eqref{privacy_constraint} and the reliability constraint \eqref{reliability_constraint}, where $D(r)$ represents the expected number of downloaded bits (over all the queries) from the $N$ databases via the answering strings $A_{1:N}^{[k]}$, i.e.,
\begin{align}
D(r)=\sum_{n=1}^N H\left(A_n^{[k]}\right). 
\end{align}
In this work, we aim to characterize the optimal normalized download cost $D^*(r)$ corresponding to every caching ratio $0 \leq r\leq 1$, where
\begin{align}
D^*(r)=\inf \left\{ \frac{D(r)}{L}: \left(D(r), L \right) \text{ is achievable}  \right\}, 
\end{align}
which is a function of the caching ratio $r$. 

\section{Main Results and Discussions}
Our first result characterizes an outer bound (achievable rate) for the normalized download cost $D^*(r)$ for general $K$, $N$, $r$.
\begin{theorem}[Outer bound]\label{Thm1}
	In the cache-aided PIR with uncoded and unknown prefetching, let $s \in \{1,2, \cdots, K-1\}$, for the caching ratio $r_s$, where
	\begin{align} \label{r_exp}
	r_s=\frac{\binom{K-2}{s-1}}{\binom{K-2}{s-1}+\sum_{i=0}^{K-1-s} \binom{K-1}{s+i}(N-1)^iN},
	\end{align}
	the optimal normalized download cost $D^*(r_s)$ is upper bounded by,
	\begin{align} \label{eq_outer} 
	D^*(r_s) \leq \bar{D}(r_s)= \frac{\sum_{i=0}^{K-1-s} \binom{K}{s+1+i}(N-1)^i N}{\binom{K-2}{s-1}+\sum_{i=0}^{K-1-s} \binom{K-1}{s+i}(N-1)^iN}
	\end{align}
	Moreover, if $r_s < r < r_{s+1}$, and $\alpha \in (0,1)$ such that $r=\alpha r_s+(1-\alpha) r_{s+1}$, then 
	\begin{align}
	D^*(r) \leq \bar{D}(r) =\alpha \bar{D}(r_s)+(1-\alpha)\bar{D}(r_{s+1})
	\end{align} 
\end{theorem}

The proof of Theorem~\ref{Thm1} can be found in Section~\ref{achievability}. Theorem~\ref{Thm1} implies that there exist $K+1$ \emph{interesting} caching ratios denoted by $r_s$, where $s \in \{1,2, \cdots, K-1\}$ in addition to $r=0$ point (no caching) and $r=1$ point (everything cached). The index $s$, which characterizes $r_s$ for these points, represents the number of cached bits that can be used within one bit of the download (if this downloaded bit uses cached bits as side information). For example, if $s=2$, this means that the user should use two of the cached bits as side information in the form of mixture of two bits if the caching ratio is $r_2$. The achievability scheme for any other caching ratio $r$ can be obtained by memory-sharing between the most adjacent interesting caching ratios that include $r$. Consequently, the outer bound is a piece-wise linear convex curve that connects the $K+1$ interesting caching ratio points including the $(0,\frac{1}{C})$ point, where $C$ is the PIR capacity without caching found in \cite{JafarPIR}, and $(1,0)$ where everything is cached. 
 
As a direct corollary for Theorem~\ref{Thm1}, we note that since the databases do not know the cached bits, the download cost is strictly smaller than the case when the databases have the full knowledge about the cached bits in \cite{tandon2017capacity}. We state and prove this in the following corollary. As a concrete example, Figure~\ref{K=5,N=2 case} shows the gain that can be achieved due to the unawareness of the databases about the cached bits.
\begin{figure}[t]
	\centering
	\epsfig{file=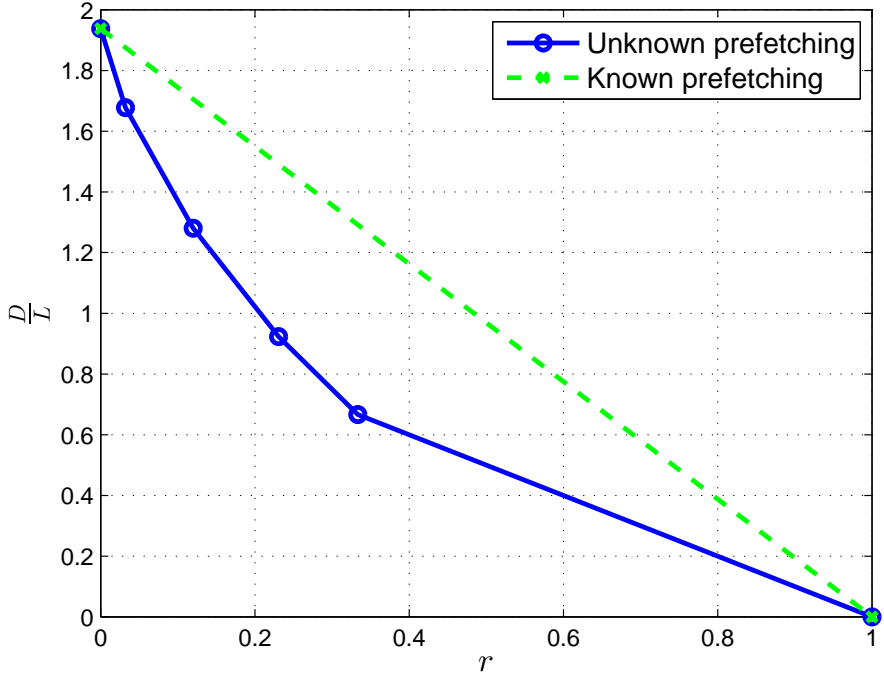,width=0.8\textwidth}
	\caption{Comparison between the optimal download cost for known prefetching \eqref{eq_pure_time_sharing} in \cite{tandon2017capacity} and the achievable download cost for unknown prefetching in \eqref{eq_outer} for $K=5$, $N=2$.}
	\label{K=5,N=2 case}
\end{figure}

\begin{corollary}[Unawareness gain]
The achievable normalized download cost $\hat{D}(r)$ in the cache-aided PIR with known prefetching \cite{tandon2017capacity} 
\begin{align} \label{eq_pure_time_sharing}
\hat{D}(r) =(1-r)\left(1+\frac{1}{N}+\dots+\frac{1}{N^{K-1}}\right)
\end{align}
is strictly larger than the achievable normalized download cost $\bar{D}(r)$ in \eqref{eq_outer}, i.e., the databases' unawareness contributes to reducing the download cost beyond the memory-sharing scheme in \cite{tandon2017capacity}. 
\end{corollary}
\begin{Proof}
For $r=0$, the achievable download cost $\bar{D}(r)$ in \eqref{eq_outer} is $\left(1+\frac{1}{N}+\dots+\frac{1}{N^{K-1}}\right)$, which is the same as \eqref{eq_pure_time_sharing}. For $r=1$, the achievable download cost $\bar{D}(r)$ in \eqref{eq_outer} is $0$, which is the same as \eqref{eq_pure_time_sharing}. To show that $\hat{D}(r)$ in \eqref{eq_pure_time_sharing} is always larger than $\bar{D}(r)$ in \eqref{eq_outer}, it suffices to show that there exists a caching ratio $r$ such that $\bar{D}(r)<\hat{D}(r)$, since the other caching ratios can be achieved by the memory-sharing scheme. Taking $s=K-1$ in \eqref{r_exp}, we have $r_{K-1}=\frac{1}{1+N}$. For $r=\frac{1}{1+N}$, we have $\bar{D}(r)=\frac{N}{1+N}$, and $\hat{D}(r)=\frac{N}{1+N}\left(1+\frac{1}{N}+\dots+\frac{1}{N^{K-1}}\right)$. 
Therefore, for $r_{K-1}$, we have $\bar{D}(r)<\hat{D}(r)$, which shows the sub-optimality of $\hat{D}(r)$ in \eqref{eq_pure_time_sharing} for the case of known prefetching.  
\end{Proof}

Our second result characterizes an inner bound (converse bound) for the normalized download cost $D^*(r)$ for general $K$, $N$, $r$.
\begin{theorem}[Inner bound]\label{Thm2}
In the cache-aided PIR with uncoded and unknown prefetching, the normalized download cost is lower bounded as,
\begin{align} \label{eq_inner}
D^*(r) \geq \tilde{D}(r)=\max_{i \in \{2, \cdots, K+1\}} \quad (1-r) \sum_{j=0}^{K+1-i} \frac{1}{N^j} - r \sum_{j=0}^{K-i} \frac{K+1-i-j}{N^j},
\end{align}
\end{theorem}

The proof of Theorem~\ref{Thm2} can be found in Section~\ref{converse}. Theorem~\ref{Thm2} implies that the inner bound is also a piece-wise linear curve, which consists of $K$ line segments with decreasing slope as $r$ increases. The points at which the curve changes its slope are given by,
\begin{align} \label{tilde_ri}
\tilde{r}_i=\frac{1}{1+N+N^2+\cdots+N^{K-i}}, \quad i=1, \cdots, K-1. 
\end{align}  
We note that $r_i$ in \eqref{r_exp} and $\tilde{r}_i$ in \eqref{tilde_ri} are the same for $i=1$ and $i=K-1$. 

As a consequence of Theorem~\ref{Thm1} and Theorem~\ref{Thm2}, we characterize the optimal download cost caching ratio tradeoff for very low and very high caching ratios in the following corollary. Here, by very low caching ratios we mean $0\leq r \leq r_1=\tilde{r}_1=\frac{1}{1+N+N^2+\cdots+N^{K-1}}$, and by very high caching ratios we mean $r_{K-2}=\frac{K-2}{(N+1)K+N^2-2N-2}\leq r\leq 1$. Note that, in the very high caching ratios, we have two segments, one in $r_{K-2}\leq r \leq r_{K-1}$ and the other in $r_{K-1} \leq r \leq 1$. Therefore, in the inner and outer bounds, each composed of $K$ line segments, the first (very low $r$) and the last two (very high $r$) segments match giving exact result. This is stated and proved in the next corollary. 

\begin{corollary} [Optimal tradeoff for very low and very high caching ratios]\label{corollarylowhigh} ~
In the cache-aided PIR with uncoded and unknown prefetching, for very low caching ratios, i.e., for $r \leq \frac{1}{ 1+N+N^2+\cdots+N^{K-1}}$, the optimal normalized download cost is given by, 
\begin{align}\label{corollary2eqn}
D^*(r)=(1-r) \left(1+\frac{1}{N}+\cdots+\frac{1}{N^{K-1}}\right)-r\left((K-1)+\frac{K-2}{N}+\cdots+\frac{1}{N^{K-2}}\right)
\end{align}
On the other hand, for very high caching ratios, i.e., for $r \geq \frac{K-2}{(N+1)K+N^2-2N-2}$, the optimal normalized download cost is given by,
\begin{align} \label{corollary2eqn2}
D^*(r) =
\left\{
\begin{array}{ll}
(1-r)\left(1+\frac{1}{N}\right)-r, &\qquad \frac{K-2}{(N+1)K+N^2-2N-2} \leq r \leq \frac{1}{1+N} \\
1-r, &\qquad\frac{1}{1+N} \leq r  \leq 1 
\end{array}
\right.
\end{align}
\end{corollary}

\begin{Proof}
First, from \eqref{r_exp} and \eqref{tilde_ri}, let us note that 
\begin{align}
r_1=\tilde{r}_1=& \frac{1}{ 1+N+N^2+\cdots+N^{K-1}},  \label{eq_tmp1} \\
r_{K-2}=&\frac{K-2}{(N+1)K+N^2-2N-2},                 \label{eq_tmp2} \\
r_{K-1}=\tilde{r}_{K-1}=&\frac{1}{1+N}.               \label{eq_tmp3}  
\end{align}
	
Then, we note from \eqref{eq_outer} that  
\begin{align}
\bar{D}(r_1) &= \frac{\sum_{i=0}^{K-2} \binom{K}{2+i}(N-1)^i N}{\binom{K-2}{0}+\sum_{i=0}^{K-2} \binom{K-1}{1+i}(N-1)^iN} \\
             &=  \frac{\frac{N}{(N-1)^{2}}\left[N^K - \sum_{i=0}^1 \binom{K}{i}(N-1)^i  \right]  }
{\binom{K-2}{0} + \frac{N}{(N-1)^{1}} \left[N^{K-1} - \sum_{i=0}^{0} \binom{K-1}{i}(N-1)^i  \right]    }   \\
&=\frac{N\left[N^K - 1 - K(N-1) \right]  }{(N-1)^{2}+N(N-1) \left[ N^{K-1}-1 \right] } \\
&=\frac{N^{K+1}-KN^2 +(K-1)N  }{N^{K+1}-N^K-N+1 }
\end{align}
Further, we note from \eqref{eq_inner}, by choosing $i=2$ and using $r=r_1$, that 
\begin{align}
\tilde{D}(r_1)&\geq(1-r_1) \sum_{j=0}^{K+1-2} \frac{1}{N^j} - r_1 \sum_{j=0}^{K-2} \frac{K-1-j}{N^j} \\
&= \left(1- \frac{N-1}{N^K-1}   \right) \frac{N^K-1}{N^K-N^{K-1}} 
-\frac{N-1}{N^K-1}  \frac{N}{1-N} \left(-K + \frac{N^K-1}{N^K-N^{K-1}}   \right) \\
&=\frac{N^K-N}{N^K-1} \frac{N^K-1}{N^K-N^{K-1}} + \frac{N}{N^K-1} \left(-K + \frac{N^K-1}{N^K-N^{K-1}}   \right) \\
&=\frac{N^K-N}{N^K-N^{K-1}} + N \left( \frac{-K}{N^K-1} + \frac{1}{N^K-N^{K-1}}    \right)  \\
&=\frac{N^{K+1}-KN^2 +(K-1)N  }{N^{K+1}-N^K-N+1 } \\
&=\bar{D}(r_1)  \label{tmp1}
\end{align}
Thus, since $\tilde{D}(r_1) \leq \bar{D}(r_1)$ by definition, \eqref{tmp1} implies $\tilde{D}(r_1) = \bar{D}(r_1)$. 

Similarly, from \eqref{eq_outer}, 
\begin{align}
\bar{D}(r_{K-2}) &= \frac{\sum_{i=0}^{1} \binom{K}{K-1+i}(N-1)^i N}{\binom{K-2}{K-3}+\sum_{i=0}^{1} \binom{K-1}{K-2+i}(N-1)^iN} \\
&= \frac{N^2 +(K-1)N}{N^2+(K-2)N+(K-2)}, 
\end{align}
and from \eqref{eq_inner} by choosing $i=K$ and using $r=r_{K-2}$, 
\begin{align}
\tilde{D}(r_{K-2})&\geq (1-r_{K-2}) \sum_{j=0}^{1} \frac{1}{N^j} -  r_{K-2}\sum_{j=0}^{0} \frac{1-j}{N^j} \\
& = \left( \frac{N^2+(K-2)N}{N^2+(K-2)N+(K-2)} \right) \left(1+\frac{1}{N} \right) - \frac{K-2}{N^2+(K-2)N+(K-2)} \\
& = \frac{N^2 +(K-1)N}{N^2+(K-2)N+(K-2)} \\
& =  \bar{D}(r_{K-2}) \label{tmp2}
\end{align}
implying $\tilde{D}(r_{K-2})=\bar{D}(r_{K-2})$. 

Finally, from \eqref{eq_outer}, 
\begin{align}
\bar{D}(r_{K-1})=\frac{N}{1+N}, 
\end{align}
and from \eqref{eq_inner} by choosing $i=K+1$ and using $r=r_{K-1}$,
\begin{align}
\tilde{D}(r_{K-1})\geq\frac{N}{1+N}=\bar{D}(r_{K-1}) \label{tmp3} 
\end{align} 
implying $\tilde{D}(r_{K-1})=\bar{D}(r_{K-1})$. 

Therefore, $\tilde{D}(r)=\bar{D}(r)$ at $r=r_1$, $r=r_{K-2}$ and $r=r_{K-1}$. We also note that $\tilde{D}(0)=\bar{D}(0)$ and $\tilde{D}(1)=\bar{D}(1)$. Since both $\bar{D}(r)$ and $\tilde{D}(r)$ are linear functions of $r$, and since $\tilde{D}(0)=\bar{D}(0)$ and $\tilde{D}(r_1)=\bar{D}(r_1)$, we have $\tilde{D}(r)=\bar{D}(r)=D^*(r)$ for $0\leq r\leq r_1$. This is the very low caching ratio region. In addition, since $\tilde{D}(r_{K-2})=\bar{D}(r_{K-2})$, $\tilde{D}(r_{K-1})=\bar{D}(r_{K-1})$ and $\tilde{D}(1)=\bar{D}(1)$, we have $\tilde{D}(r)=\bar{D}(r)=D^*(r)$ for $r_{K-2}\leq r\leq 1$. This is the very high caching ratio region. 
\end{Proof}

As an example, the case of $K=4$, $N=2$ is shown in Figure~\ref{K=4,N=2 case}. In this case, $r_1=\tilde{r}_1=\frac{1}{15}$, $r_{K-2}=\frac{1}{5}$, and $r_{K-1}=\tilde{r}_{K-1}=\frac{1}{3}$. Therefore, we have exact results for $0\leq r \leq \frac{1}{15}$ (very low caching ratios) and $\frac{1}{5}\leq r \leq 1$ (very high caching ratios). We have a gap between the achievability and the converse for medium caching ratios in $\frac{1}{15}\leq r \leq \frac{1}{5}$. More specifically, line segments connecting $(0,\frac{15}{8})$ and $(\frac{1}{15},\frac{22}{15})$; connecting $(\frac{1}{5},1)$ and $(\frac{1}{3},\frac{2}{3})$; and connecting $(\frac{1}{3},\frac{2}{3})$ and $(1,0)$ are tight. 
\begin{figure}[t]
	\centering
	\epsfig{file=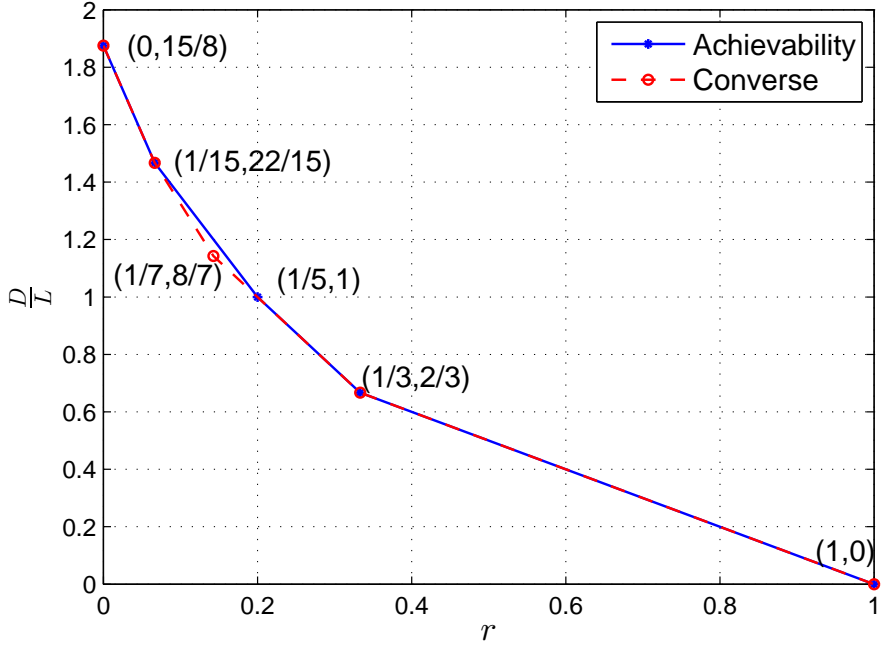,width=0.8\textwidth}
	\caption{Inner and outer bounds for $K=4$, $N=2$.}
	\label{K=4,N=2 case}
\end{figure}

\begin{figure}[t]
	\centering
	\epsfig{file=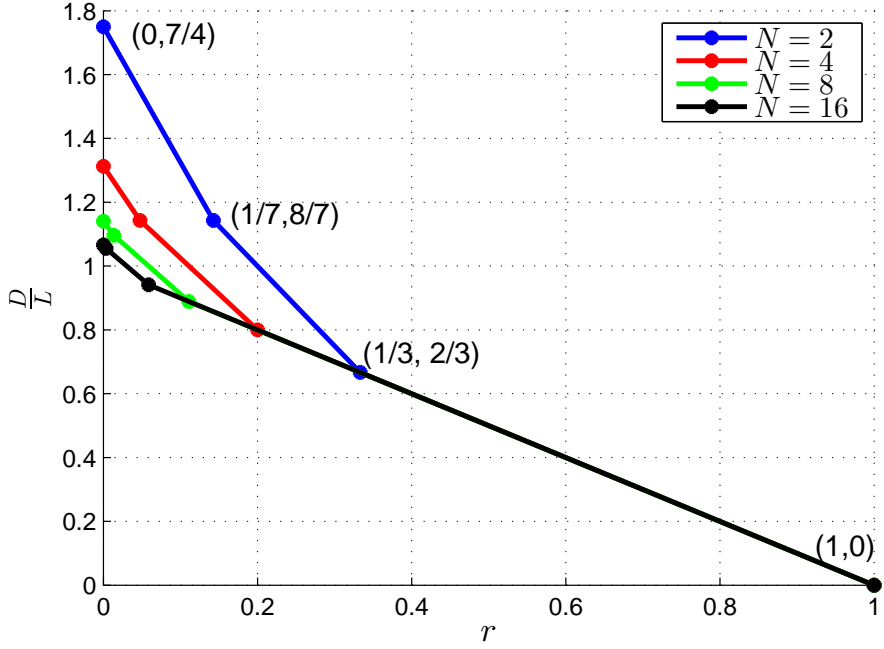,width=0.8\textwidth}
	\caption{Optimal download cost caching ratio tradeoff for the case of $K=3$ messages.}
	\label{K=3_tradeoff}
\end{figure}

Finally, we characterize the exact tradeoff curve for any $N$, $r$ for the special case of $K=3$ in the following corollary.
\begin{corollary}[Optimal tradeoff for $K=3$]\label{corollaryK=3}
	In the cache-aided PIR with uncoded and unknown prefetching with $K=3$ messages, the optimal download cost caching ratio tradeoff is given explicitly as (see Figure~\ref{K=3_tradeoff}),
	\begin{equation} \label{tmp4}
	D^*(r) =
	\left\{
	\begin{array}{ll}
	(1-r)\left(1+\frac{1}{N}+\frac{1}{N^2}\right)-r\left(2+\frac{1}{N}\right), &\qquad 0 \leq r \leq \frac{1}{1+N+N^2} \\
	(1-r)\left(1+\frac{1}{N}\right)-r, &\qquad\frac{1}{1+N+N^2} \leq r \leq \frac{1}{1+N} \\
	1-r, &\qquad\frac{1}{1+N} \leq r  \leq 1 
	\end{array}
	\right.
	\end{equation}
\end{corollary}
\begin{Proof}
The proof follows from the proof of Corollary~\ref{corollarylowhigh}. Note that in this case, from \eqref{eq_tmp1} and \eqref{eq_tmp2}, $r_1=r_{K-2}=\frac{1}{1+N+N^2}$; and from \eqref{eq_tmp3}, $r_2=r_{K-1}=\frac{1}{1+N}$. Thus, we have a tight result for $0\leq r \leq r_1=\frac{1}{1+N+N^2}$ (very low caching ratios) and a tight result for $r_{K-2}=r_1=\frac{1}{1+N+N^2}\leq r \leq 1$, i.e., a tight result for all $0\leq r \leq 1$. We have three segments in this case: $[0, r_1]$, $[r_1, r_2]$ and $[r_2,1]$ with three different line expressions for the exact result as given in 
\eqref{r_exp}-\eqref{eq_outer} and written explicitly in \eqref{tmp4}. 
\end{Proof}

\section{Achievability Proof}\label{achievability}
Our achievability scheme is based on the PIR schemes in \cite{JafarPIR,tandon2017capacity}. Similar to \cite{JafarPIR}, we apply the following three principles recursively: 1) database symmetry, 2) message symmetry within each database, and 3) exploiting undesired messages as side information. Different from \cite{JafarPIR}, we start the PIR scheme from the third principle due to the availability of pre-existing side information as a result of uncoded prefetching. These cached bits can be exploited right away as side information without compromising the privacy constraint as the databases do not know them. We begin the discussion by presenting the case of $K=3$, $N=2$ as a motivating example to illustrate the main ideas of our achievability scheme. 

\subsection{Motivating Example: The Optimal Tradeoff Curve for $K=3$ Messages, $N=2$ Databases}
In this example, we show the achievability for $K=3$, $N=2$. We know from Corollary~\ref{corollaryK=3} that the inner and the outer bounds match for this case. The optimal download cost caching ratio tradeoff is shown in Figure~\ref{K=3_tradeoff}. We note that there are $4$ corner points. Two fo them are degenerate, corresponding to $r=0$, $r=1$ caching ratios. For $r=0$, the user has no cached bits and is forced to apply the achievable scheme in \cite{JafarPIR} that achieves $\bar{D}(0)=\frac{7}{4}=\frac{1}{C}$. For $r=1$, the user has already cached the entire desired file and does not download any extra bits from the databases, i.e., $\bar{D}(1)=0$. We have two other corner points, corresponding to $r_1=\frac{1}{1+N+N^2}=\frac{K-2}{(N+1)K+N^2-2N-2}=\frac{1}{7}$, and $r_2=\frac{1}{1+N}=\frac{1}{3}$. In the sequel, we show the achievability of these two corner points.  

\subsubsection{Caching Ratio $r_1=\frac{1}{7}$} \label{r=1/7}

Let $s$ be the number of cached bits that are mixed together to form side information equation. The first corner point corresponds to $s=1$. This means that the user exploits every bit in the cache individually as a side information. Using the notation in \cite{MPIRjournal}, we can say that the user starts downloading from round 2 that sums bits from very two messages together. We next show how $s=1$ suffices to achieve $r_1=\frac{1}{7}$, $\bar{D}(\frac{1}{7})=\frac{8}{7}$ for $K=3$, $N=2$; see Figure~\ref{K=3_tradeoff}. 

We use $a_i$, $b_i$, and $c_i$ to denote the bits of messages $W_1$, $W_2$ and $W_3$, respectively. We assume that the user wants to retrieve message $W_1$ privately without loss of generality. We initialize the process by permuting the indices of messages $W_1, W_2, W_3$ randomly and independently. The steps of the retrieval can be followed in Table~\ref{K=3 r=1/7}. The user has already cached one bit from each message, i.e., $a_1$, $b_1$, $c_1$ as denoted by $Z$ in Table~\ref{K=3 r=1/7}. We start from the third principle by exploiting each bit in the cache as an individual side information. The user downloads $a_2+b_1$ and $a_3+c_1$ from the first database (DB1). Then, we apply the first principle, and the user downloads $a_4+b_1$ and $a_5+c_1$ from the second database (DB2) to satisfy the database symmetry. Next, we apply the second principle to ensure the message symmetry within the queries. The user downloads $b_2+c_2$ from DB1, and $b_3+c_3$ from DB2. At this point, all side information corresponding to the cached bits have been exploited. Next, we apply the third principle, since undesired message mixes are available in the form of $b_2+c_2$ and $b_3+c_3$. The user downloads $a_6+b_3+c_3$ from DB1. Finally, we apply the first principle of database symmetry, and the user downloads $a_7+b_2+c_2$ from DB2. Now, the iterations stop, since all the undesired side information is used and the symmetry across databases and symmetry within the queries is attained. We summarize the process in the query table in Table~\ref{K=3 r=1/7}. 

\begin{table}[h]
	\centering
	\caption{Query table for $K=3$, $N=2$, $r_1=\frac{1}{7}$}
	\label{K=3 r=1/7}
	\begin{tabular}{ccc}
		\hline
		\multicolumn{1}{|c|}{$s$}                    & \multicolumn{1}{c|}{DB1}           & \multicolumn{1}{c|}{DB2}           \\ \hline
		\multicolumn{1}{|c|}{\multirow{2}{*}{\rotatebox[origin=c]{90}{\parbox[c]{1cm}{\centering $s=1$}}}} & \multicolumn{1}{c|}{$a_2+b_1$}     & \multicolumn{1}{c|}{$a_4+b_1$}     \\ \cline{2-3} 
		\multicolumn{1}{|c|}{}                       & \multicolumn{1}{c|}{$a_3+c_1$}     & \multicolumn{1}{c|}{$a_5+c_1$}     \\ \hline
		\multicolumn{1}{|c|}{\multirow{2}{*}{}}      & \multicolumn{1}{c|}{$b_2+c_2$}     & \multicolumn{1}{c|}{$b_3+c_3$}     \\ \cline{2-3} 
		\multicolumn{1}{|c|}{}                       & \multicolumn{1}{c|}{$a_6+b_3+c_3$} & \multicolumn{1}{c|}{$a_7+b_2+c_2$} \\ \hline
		&                                    &                                    \\ \cline{2-3} 
		\multicolumn{1}{c|}{}                        & \multicolumn{2}{c|}{$Z=(a_1,b_1,c_1)$}                                  \\ \cline{2-3} 
	\end{tabular}
\end{table}

Since the databases do not know the local cache memory $Z$, and for each database, the user's queries are symmetric across messages, the privacy constraint \eqref{privacy_constraint} is satisfied. The decodability can be easily checked as the user can cancel out $b_1,c_1$ which it has previously cached, and also cancel $b_2+c_2$ and $b_3+c_3$ which are previously downloaded, to obtain $a_2,\cdots, a_7$. Since $a_1$ is already cached, the user has $a_1, \cdots, a_7$. Here, $L=7$ and the user has cached $1$ bit from each message. There are total of $8$ downloads. Hence $r=\frac{1}{7}$, and $\bar{D}(\frac{1}{7})=\frac{8}{7}$. 

\subsubsection{Caching Ratio $r_2=\frac{1}{3}$} \label{r=1/3}
For the second non-degenerate corner point, we have $s=2$. This means that each 2 bits from the cache are mixed together to form a side information equation. We next show how $s=2$ suffices to achieve $r_2=\frac{1}{3}$, $\bar{D}(\frac{1}{3})=\frac{2}{3}$ for $K=3$, $N=2$; see Figure~\ref{K=3_tradeoff}. 

Let $[a_1, a_2, a_3]$, $[b_1, b_2, b_3]$, and $[c_1, c_2, c_3]$ denote a random permutation of the $3$ bits of messages $W_1$, $W_2$ and $W_3$, respectively. Suppose the user caches $a_1, b_1, c_1$ in advance and wants to retrieve message $W_1$ privately. We start from the third principle. The user downloads $a_2+b_1+c_1$ from the first database (DB1). Then, we apply the first principle, and the user downloads $a_3+b_1+c_1$ from the second database (DB2). Now, the iterations stop, since all the undesired side information is used and the symmetry across databases and messages is attained. We summarize the process in the query table in Table~\ref{K=3 r=1/3}. In this case $L=3$, hence $r=\frac{1}{3}$, and the normalized download cost is $\bar{D}(\frac{1}{3})=\frac{2}{3}$. 

\begin{table}[h]
	\centering
	\caption{Query table for $K=3$, $N=2$, $r_2=\frac{1}{3}$}
	\label{K=3 r=1/3}
	\begin{tabular}{ccc}
		\hline
		\multicolumn{1}{|c|}{$s$}   & \multicolumn{1}{c|}{DB1}              & \multicolumn{1}{c|}{DB2}           \\ \hline
		\multicolumn{1}{|c|}{$s=2$} & \multicolumn{1}{c|}{$a_2+b_1+c_1$} & \multicolumn{1}{c|}{$a_3+b_1+c_1$} \\ \hline
		&                                       &                                    \\ \cline{2-3} 
		\multicolumn{1}{c|}{}       & \multicolumn{2}{c|}{$Z=(a_1,b_1,c_1)$}                                     \\ \cline{2-3} 
	\end{tabular}
\end{table}

\subsubsection{Caching Ratio $r=\frac{1}{5}$}
So far, we have characterized all the corner points by varying $s=1,2$ and achieved the points corresponding to caching ratios $r_s$ in addition to the degenerate caching ratios $r=0$ and $r=1$; see Figure~\ref{K=3_tradeoff}. An achievable scheme for any other caching ratio can be obtained by memory-sharing between the two nearest corner points. As an example, we next consider the caching ratio $r=\frac{1}{5}$. 

The achievability scheme for this case is a combination of the achievability schemes in Sections~\ref{r=1/7} and \ref{r=1/3}. Observe that by choosing $L=10$, the achievable schemes in Sections~\ref{r=1/7} and \ref{r=1/3}  can be concatenated to achieve the caching ratio $r=\frac{1}{5}$. In this case, the user caches $a_1, a_2, b_1, b_2 ,c_1, c_2$ and wants to retrieve message $W_1$ privately. For cached bits $a_1,b_1,c_1$, we apply the same process as in Section~\ref{r=1/7}, i.e., we use $s=1$ and use every cached bit as individual side information equation. For cached bits $a_2,b_2,c_2$, we apply the same process as in Section~\ref{r=1/3}, and choose $s=2$, which implies that we use the mixture of two cached bits as a side information equation. We summarize the process in the query table in Table~\ref{K=3 r=1/5}. 

\begin{table}[h]
	\centering
	\caption{Query table for $K=3$, $N=2$, $r=\frac{1}{5}$}
	\label{K=3 r=1/5}
	\begin{tabular}{ccc}
		\hline
		\multicolumn{1}{|c|}{$s$}                    & \multicolumn{1}{c|}{DB1}           & \multicolumn{1}{c|}{DB2}              \\ \hline
		\multicolumn{1}{|c|}{\multirow{2}{*}{$s=1$}} & \multicolumn{1}{c|}{$a_3+b_1$}     & \multicolumn{1}{c|}{$a_5+b_1$}        \\ \cline{2-3} 
		\multicolumn{1}{|c|}{}                       & \multicolumn{1}{c|}{$a_4+c_1$}     & \multicolumn{1}{c|}{$a_6+c_1$}        \\ \hline
		\multicolumn{1}{|c|}{\multirow{2}{*}{}}      & \multicolumn{1}{c|}{$b_3+c_3$}     & \multicolumn{1}{c|}{$b_4+c_4$}        \\ \cline{2-3} 
		\multicolumn{1}{|c|}{}                       & \multicolumn{1}{c|}{$a_7+b_4+c_4$} & \multicolumn{1}{c|}{$a_8+b_3+c_3$}    \\ \hline
		\multicolumn{1}{|c|}{$s=2$}                  & \multicolumn{1}{c|}{$a_9+b_2+c_2$} & \multicolumn{1}{c|}{$a_{10}+b_2+c_2$} \\ \hline
		\multicolumn{1}{l}{}                         & \multicolumn{1}{l}{}               & \multicolumn{1}{l}{}                  \\ \cline{2-3} 
		\multicolumn{1}{l|}{}                        & \multicolumn{2}{c|}{$Z=(a_1, a_2, b_1, b_2 ,c_1, c_2)$}                    \\ \cline{2-3} 
	\end{tabular}
\end{table}

Here, we have $L=10$, therefore $r=\frac{1}{5}$, and $\bar{D}(\frac{1}{5})=\frac{10}{10}=1$. In fact, by applying \cite[Lemma~1]{tandon2017capacity} and taking $\alpha=\frac{7}{10}$, we can show that the normalized download cost of this example can be obtained from the download costs obtained in Sections~\ref{r=1/7} and \ref{r=1/3}, as $\bar{D}(\frac{1}{5})=\bar{D}(\frac{1}{7} \cdot \frac{7}{10} + \frac{1}{3}\cdot \frac{3}{10} )=\frac{7}{10}\bar{D}(\frac{1}{7})+\frac{3}{10}\bar{D}(\frac{1}{3})=\frac{7}{10}\cdot\frac{8}{7}+\frac{3}{10}\cdot\frac{2}{3}=1$. 

\subsection{Achievable Scheme for the Corner Points for Arbitrary $K$, $N$}
For fixed $N$ and $K$, there are $K-2$ non-degenerate corner points (in addition to degenerate caching ratios $r=0$, $r=1$). The caching ratios corresponding to these non-degenerate corner points are indexed by $s$, which enumerate the number of cached bits that are involved in the side information mixture. Hence, $r_s$ is given by  
\begin{align} \label{r_exact}
r_s=\frac{\binom{K-2}{s-1}}{\binom{K-2}{s-1}+\sum_{i=0}^{K-1-s} \binom{K-1}{s+i}(N-1)^iN},
\end{align}
where $s\in\{1,2,\dots, K-1\}$. We choose the length of the message to be $L(s)$ for the corner point indexed by $s$,  where
\begin{align} \label{L_eq}
L(s)=\binom{K-2}{s-1}+\sum_{i=0}^{K-1-s} \binom{K-1}{s+i}(N-1)^iN
\end{align} 
bits per message. The details of the achievable scheme are as follows:
\begin{enumerate}
	\item \emph{Initialization:} The user permutes each message randomly and independently. The user caches randomly and privately $\binom{K-2}{s-1}$ bits from each message. Set the round index to $i=s+1$, where the $i$th round involves downloading sums of every $i$ combinations of the $K$ messages.
	\item \emph{Exploiting side information:} If $i=s+1$, the user mixes $s$ bits from the cache bits to form one side information equation. Each side information equation is added to one bit from the uncached portion of the desired message. Therefore, the user downloads $\binom{K-1}{s}$ equations in the form of a desired bit added to a mixture of $s$ cached bits from other messages. On the other hand, if $i>s+1$, the user exploits the $\binom{K-1}{i-1}(N-1)^{i-s-1}$ side information equations generated from the remaining $(N-1)$ databases in the $(i-1)$th round.
	\item \emph{Symmetry across databases:} The user downloads the same number of equations with the same structure as in step $2$ from every database. Consequently, the user downloads $\binom{K-1}{i-1}(N-1)^{i-s-1}$ bits from every database, which are done either using the cached bits as side information if $i=s+1$, or the side information generated in the $(i-1)$th round if $i>s+1$.
	\item \emph{Message symmetry:} To satisfy the privacy constraint, the user should download equal amount of bits from all other messages. Therefore, the user downloads $\binom{K-1}{i}(N-1)^{i-s-1}$ undesired equations from each database in the form of sum of $i$ bits from the uncached portion of the undesired messages.
	\item \emph{Repeat} steps 2, 3, 4 after setting $i=i+1$ until $i=K$.
	\item \emph{Shuffling the order of queries:} By shuffling
	the order of queries uniformly, all possible queries can be made equally likely regardless of the message index. This guarantees the privacy.
\end{enumerate}

\subsubsection{Decodability, Privacy, and the Achievable Normalized Download Cost}
\paragraph{Decodability:} It is clear that the side information in each round is either constructed from the cached bits (if $i=s+1$) or obtained from the remaining $(N-1)$ databases in the $(i-1)$th round. Consequently, the user can cancel out these side information bits in order to decode the uncached portion of the desired message (the remaining $L(1-r)$ bits).

\paragraph{Privacy:} The randomized mapping of the cached and the uncached portions of the messages and the randomization of the order of queries guarantees privacy as in \cite{JafarPIR}.

\paragraph{Normalized Download Cost:} We now calculate the total number of downloaded bits for the caching ratio $r$ in \eqref{r_exact}. First, we exploit $s$ bits of side information. Therefore, each download is a sum of $s+1$ bits. Since the second principle enforces symmetry across $K$ messages, we download $\binom{K}{s+1}$ bits from a database. Due to the first principle enforcing symmetry across databases, in total, we download $\binom{K}{s+1}N$ bits. Since we utilize $s$ bits of side information of undesired messages for each download, for each undesired message we use $\frac{\binom{K-1}{s} s}{K-1}=\binom{K-2}{s-1}$ bits, which is the amount of bits we cached in advance for each message. Next, each download is a sum of $s+2$ bits since the available side information is in the form of sums of $s+1$ bits. Due to message symmetry and $(N-1)$ available side information from other $(N-1)$ databases, we download $\binom{K}{s+2}(N-1)$ bits from each database. Due to the first principle enforcing symmetry across databases, in total, we download $\binom{K}{s+2}(N-1)N$ bits. Next, each download is sum of $s+3$ bits since the available side information is in the form of sums of $s+2$ bits. Note that in the previous iteration, each database provides $(N-1)$ sets of side information, and each database exploits the side information from the other $(N-1)$ databases. Therefore, we download $\binom{K}{s+3}(N-1)^2$ bits from each database. Due to the first principle enforcing symmetry across databases, in total, we download $\binom{K}{s+3}(N-1)^2N$ bits. By continuing in this manner, the total number of downloaded bits is,
\begin{align} 
D(r_s)=\sum_{i=0}^{K-1-s} \binom{K}{s+1+i}(N-1)^iN. 
\end{align} 

Now, we calculate the number of desired bits we have downloaded in this process. At the beginning of the iteration, each download is a sum of $s+1$ bits. If the download includes a desired bit, the other $s$ bits are from the local cache memory. Therefore, we download $\binom{K-1}{s}$ desired bits from each database, and thus we download a total of $\binom{K-1}{s}N$ desired bits. Next, each download is sum of $s+2$ bits. If the download includes a desired bit, the other $s+1$ bits are from the side information of undesired bits. For each database, there are $(N-1)$ sets of side information obtained from the previous iteration with one set from each database. Therefore, we download $\binom{K-1}{s+1}(N-1)$ bits from each database, and thus we download a total of $\binom{K-1}{s+1}(N-1)N$ desired bits. Next, each download is sum of $s+3$ bits. If the download includes a desired bit, the other $s+2$ bits are from the side information of undesired bits. For each database, there are $(N-1)^2$ sets of side information obtained from the previous iteration with $(N-1)$ sets from one database. Therefore, we download $\binom{K-1}{s+2}(N-1)^2N$ desired bits, and thus the number of desired bits we downloaded is $L(s)-\binom{K-2}{s-1}$, where $L(s)$ is given in \eqref{L_eq}. Finally, the normalized download cost is,  
\begin{align} \label{D(r_s)}
\bar{D}(r_s)=\frac{D(r_s)}{L(s)}=\frac{\sum_{i=0}^{K-1-s} \binom{K}{s+1+i}(N-1)^i N}{\binom{K-2}{s-1}+\sum_{i=0}^{K-1-s} \binom{K-1}{s}(N-1)^iN}.
\end{align}

\subsection{Achievable Scheme for Non-Corner Points for Arbitrary $K$, $N$}

For caching ratios $r$ which are not exactly equal to \eqref{r_exact} for some $s$, we first find an $s$ such that $r_s<r<r_{s+1}$, and combine the achievability schemes of $r_s$ and $r_{s+1}$. Then, we can write the achievable normalized download cost as a convex combination of $\bar{D}(r_s)$ and $\bar{D}(r_{s+1})$ using \cite[Lemma~1]{tandon2017capacity} as follows,
\begin{align}
\bar{D}(r) = \alpha \bar{D}(r_s)+(1-\alpha) \bar{D}(r_{s+1}), 
\end{align}
where $r=\alpha r_s + (1-\alpha) r_{s+1}$ and $r_s$ is defined in \eqref{r_exact}, and $\bar{D}(r)$ is given in \eqref{D(r_s)}.

\section{Converse Proof}\label{converse}
In this section, we derive an inner bound for the cache-aided PIR with uncoded and unknown prefetching. The inner bound is tight in general for very high and very low caching ratios, and in particular, the inner bound is tight everywhere for $K=3$. We extend the techniques presented in \cite{JafarPIR,tandon2017capacity} to our problem. We first need the following lemma, which characterizes a lower bound on the length of the undesired portion of the answer strings as a consequence of the privacy constraint. 
\begin{lemma}[Interference lower bound]\label{lemma_converse1}
	For the cache-aided PIR with unknown and uncoded prefetching, the interference from undesired messages within the answer strings $D(r)-L(1-r)$ is lower bounded by, 	
	\begin{align}
	D(r) -L(1-r) + o(L) \geq I\left(W_{k:K}; \mathbb{H}, Q_{1:N}^{[k-1]}, A_{1:N}^{[k-1]}|W_{1:k-1}, Z \right) \label{eq_L1}
	\end{align}
	for all $k\in \{2,\dots,K\}$.
\end{lemma}

If the privacy constraint is absent, the user downloads only $L(1-r)$ bits in order to decode the desired message, however, when the privacy constraint is present, it should download $D(r)$. The difference $D(r)-L(1-r)$ corresponds to the undesired portion of the answer strings. Lemma~\ref{lemma_converse1} shows that this portion is lower bounded  by the mutual information between the answer strings and the messages $W_{k:K}$ after knowing the first $W_{1:k-1}$ messages and the cached bits. Lemma~\ref{lemma_converse1} provides $K-1$ lower bounds on $D(r)-L(1-r)$ by changing the index $k$ from $2$ to $K$. Each of these $K-1$ bounds contributes a different line segment for the final inner bound. Note that Lemma~\ref{lemma_converse1} is an extension to \cite[Lemma~5]{JafarPIR} if $k=2$, $r=0$.      
\begin{Proof}
	We start with the right hand side of \eqref{eq_L1},	
	\begin{align}
	&I\left(W_{k:K}; \mathbb{H}, Q_{1:N}^{[k-1]}, A_{1:N}^{[k-1]} |W_{1:k-1}, Z \right) \notag \\
	&\qquad = I\left(W_{k:K}; \mathbb{H}, Q_{1:N}^{[k-1]}, A_{1:N}^{[k-1]}, W_{k-1} |W_{1:k-2}, Z \right) 
	- I \left(W_{k:K}; W_{k-1}|W_{1:k-2}, Z  \right) \label{eq_lemma_1}
	\end{align}	
	For the first term on the right hand side of \eqref{eq_lemma_1}, we have 
	\begin{align}
	& I\left(W_{k:K}; \mathbb{H}, Q_{1:N}^{[k-1]}, A_{1:N}^{[k-1]}, W_{k-1} |W_{1:k-2}, Z \right)  \notag \\
	&\qquad= I\left(W_{k:K}; \mathbb{H}, Q_{1:N}^{[k-1]}, A_{1:N}^{[k-1]} |W_{1:k-2}, Z \right)    + I \left( W_{k:K}; W_{k-1}|\mathbb{H}, Q_{1:N}^{[k-1]}, A_{1:N}^{[k-1]},W_{1:k-2}, Z \right)  \\
	&\label{eq_ILB_1} \qquad\stackrel{\eqref{reliability_constraint}}{=}
	I\left(W_{k:K}; \mathbb{H}, Q_{1:N}^{[k-1]}, A_{1:N}^{[k-1]} |W_{1:k-2}, Z \right)  + o(L) \\
	&\label{eq_ILB_2} \:\,\quad\stackrel{\eqref{independency},\eqref{query_indep}}{=} 
	I\left(W_{k:K}; A_{1:N}^{[k-1]} |W_{1:k-2}, Z, \mathbb{H}, Q_{1:N}^{[k-1]} \right) + o(L) \\
	&\qquad= H\left(A_{1:N}^{[k-1]} |W_{1:k-2}, Z, \mathbb{H}, Q_{1:N}^{[k-1]} \right) 
	- H\left( A_{1:N}^{[k-1]} |W_{1:k-2}, Z, \mathbb{H}, Q_{1:N}^{[k-1]}, W_{k:K} \right) + o(L) \\
	&\label{eq_ILB_3}\qquad\stackrel{\eqref{reliability_constraint}}{=}   
	H\left(A_{1:N}^{[k-1]} |W_{1:k-2}, Z, \mathbb{H}, Q_{1:N}^{[k-1]} \right)   \notag \\
	&\qquad\quad- H\left( W_{k-1}, A_{1:N}^{[k-1]} |W_{1:k-2}, Z, \mathbb{H}, Q_{1:N}^{[k-1]}, W_{k:K} \right) + o(L) \\
	&\label{eq_ILB_4}\qquad\leq H\left(A_{1:N}^{[k-1]} |W_{1:k-2}, Z, \mathbb{H}, Q_{1:N}^{[k-1]} \right)  
	- H\left( W_{k-1} |W_{1:k-2}, Z, \mathbb{H}, Q_{1:N}^{[k-1]}, W_{k:K} \right) + o(L) \\
	&\label{eq_ILB_5}\:\,\quad \stackrel{\eqref{independency},\eqref{query_indep}}{=} 
	H\left(A_{1:N}^{[k-1]} |W_{1:k-2}, Z, \mathbb{H}, Q_{1:N}^{[k-1]} \right)  
	- H\left( W_{k-1}|Z\right) + o(L) \\
	&\label{eq_ILB_6}\qquad = H\left(A_{1:N}^{[k-1]} |W_{1:k-2}, Z, \mathbb{H}, Q_{1:N}^{[k-1]} \right)   - L(1-r)+ o(L)  \\
	&\qquad \leq D(r) - L(1-r)+ o(L) \label{eq_l1_1}
	\end{align}
	where \eqref{eq_ILB_1}, \eqref{eq_ILB_3} follow from the reliability constraint of $W_{k-1}$, \eqref{eq_ILB_2} follows from the independence between the queries $Q_{1:N}^{[k-1]}$ and the messages $W_{k:K}$, \eqref{eq_ILB_4} follows from the non-negativity of the entropy function, \eqref{eq_ILB_5} is due to the fact that $W_{k-1}$ is statistically independent of $(W_{1:k-2}, W_{k:K},  \mathbb{H},Q_{1:N}^{[k-1]})$, \eqref{eq_ILB_6} follows from the uncoded nature of the cache, and \eqref{eq_l1_1} follows from conditioning reduces entropy.  
	
	For the second term on the right hand side of \eqref{eq_lemma_1}, we have
	\begin{align}
	I \left(W_{k:K}; W_{k-1}|W_{1:k-2}, Z  \right)&=H\left(W_{k-1}|W_{1:k-2}, Z\right) -H\left(W_{k-1}|W_{1:k-2}, W_{k:K},Z\right) \\
	&=\left(L-Lr \right)-\left(L-Lr \right)          \\
	&=0   \label{eq_l1_2}
	\end{align}
	
	Combining \eqref{eq_lemma_1}, \eqref{eq_l1_1}, and \eqref{eq_l1_2} yields \eqref{eq_L1}. 
\end{Proof}

In the following lemma, we prove an inductive relation for the mutual information term on the right hand side of \eqref{eq_L1}.
\begin{lemma}[Induction lemma]\label{lemma_converse2}
	For all $k\in \{2,\dots,K\}$, the mutual information term in Lemma~\ref{lemma_converse1} can be inductively lower bounded as,
	\begin{align} \label{eq_L2}
	&I\left( W_{k:K} ; \mathbb{H}, Q_{1:N}^{[k-1]}, A_{1:N}^{[k-1]} | W_{1:k-1}, Z \right)  \notag \\
	&\quad\quad\quad \geq \frac{1}{N}  I\left(W_{k+1:K};\mathbb{H}, Q_{1:N}^{[k]}, A_{1:N}^{[k]}|W_{1:k},  Z  \right)  +\frac{L(1-r) - o(L)}{N}  -(K-k+1)Lr.
	\end{align}
\end{lemma}

Lemma~\ref{lemma_converse2} relates the mutual information between $W_{k:K}$ and the answer strings to the same mutual information term with $W_{k+1:K}$, i.e., it shifts the term by one message. Since the two terms have the same structure, Lemma~\ref{lemma_converse2} constructs an inductive relation. We obtain an explicit lower bound for $I\left(W_{k:K};\mathbb{H}, Q_{1:N}^{[k-1]},A_{1:N}^{[k-1]}|W_{1:k-1},Z\right)$ by applying this lemma $K-k+1$ times, and therefore characterize an explicit lower bound on $D(r)-L(1-r)$. We do this in Lemma~\ref{Lemma_summary} by combining Lemma~\ref{lemma_converse1} and Lemma~\ref{lemma_converse2}.  Lemma~\ref{lemma_converse2} reduces to \cite[Lemma~6]{JafarPIR} if $r=0$.
\begin{Proof}
	We start with the left hand side of \eqref{eq_L2},
	\begin{align}
	&I\left( W_{k:K} ; \mathbb{H}, Q_{1:N}^{[k-1]}, A_{1:N}^{[k-1]} | W_{1:k-1}, Z \right)       \notag \\
	&\qquad  =  I\left(W_{k:K} ; \mathbb{H}, Q_{1:N}^{[k-1]}, A_{1:N}^{[k-1]}, Z | W_{1:k-1}  \right)  - I(W_{k:K};Z|W_{1:k-1})    \\
	&\qquad = I\left(W_{k:K} ; Q_{1:N}^{[k-1]}, A_{1:N}^{[k-1]}| W_{1:k-1}  \right)  +I\left(W_{k:K};\mathbb{H}, Z|W_{1:k-1}, Q_{1:N}^{[k-1]}, A_{1:N}^{[k-1]} \right) \notag \\
	&\qquad\quad- I(W_{k:K};Z|W_{1:k-1})              \\      
	&\qquad \geq I\left(W_{k:K} ; Q_{1:N}^{[k-1]}, A_{1:N}^{[k-1]}| W_{1:k-1}  \right) - I(W_{k:K};Z|W_{1:k-1}) \label{eq_L2_a}
	\end{align}
	where \eqref{eq_L2_a} follows from the non-negativity of mutual information.  
	
	For the first term in \eqref{eq_L2_a}, we have
	\begin{align}
	& N  I\left(W_{k:K} ; Q_{1:N}^{[k-1]}, A_{1:N}^{[k-1]}| W_{1:k-1}  \right) \notag \\
	& \label{eq_IL_1}\qquad \geq  \sum_{n=1}^N I\left(W_{k:K} ; Q_n^{[k-1]}, A_n^{[k-1]}| W_{1:k-1}  \right) \\
	& \label{eq_IL_2}\qquad \stackrel{\eqref{privacy_constraint}}{=}    \sum_{n=1}^N I\left(W_{k:K} ; Q_n^{[k]}, A_n^{[k]}| W_{1:k-1}  \right) \\ 
	& \label{eq_IL_444}\qquad \stackrel{\eqref{independency}}{=}  \sum_{n=1}^N I\left(W_{k:K} ; A_n^{[k]}| W_{1:k-1}, Q_n^{[k]}   \right) \\
	& \label{eq_IL_44}\qquad \stackrel{\eqref{answer_constraint}}{=}    \sum_{n=1}^N H\left(A_n^{[k]}| W_{1:k-1}, Q_n^{[k]}\right) \\
	& \label{eq_IL_3}\qquad \geq  \sum_{n=1}^N H\left(A_n^{[k]}| W_{1:k-1}, \mathbb{H}, Q_{1:N}^{[k]}, A_{1:n-1}^{[k]}, Z\right) \\ 
	& \label{eq_IL_4}\qquad \stackrel{\eqref{answer_constraint}}{=}\sum_{n=1}^N I\left(W_{k:K};A_n^{[k]}| W_{1:k-1},\mathbb{H}, Q_{1:N}^{[k]}, A_{1:n-1}^{[k]}, Z   \right)  \\
	&\label{eq_IL_5}  \qquad = I\left(W_{k:K}; A_{1:N}^{[k]} | W_{1:k-1}, \mathbb{H},Q_{1:N}^{[k]}, Z \right) \\
	&\label{eq_IL_6} \qquad \stackrel{\eqref{independency}}{=}   I\left(W_{k:K}; \mathbb{H}, Q_{1:N}^{[k]}, A_{1:N}^{[k]} | W_{1:k-1},  Z \right) \\
	& \label{eq_IL_7} \qquad \stackrel{\eqref{reliability_constraint}}{=} I\left(W_{k:K}; W_k,\mathbb{H}, Q_{1:N}^{[k]}, A_{1:N}^{[k]} | W_{1:k-1},  Z \right) - o(L) \\
	& \qquad = I\left(W_{k:K}; \mathbb{H}, Q_{1:N}^{[k]}, A_{1:N}^{[k]}|W_{1:k},  Z  \right) +I\left(W_{k:K}; W_k| W_{1:k-1},  Z \right)  - o(L) \\
	& \label{eq_IL_8}\qquad = I\left(W_{k:K}; \mathbb{H}, Q_{1:N}^{[k]}, A_{1:N}^{[k]}|W_{1:k},  Z  \right) + L(1-r) - o(L) \\
	& \qquad = I\left(W_{k+1:K};\mathbb{H}, Q_{1:N}^{[k]}, A_{1:N}^{[k]}|W_{1:k},  Z  \right) + L(1-r) - o(L) \label{eq_L2_b}
	\end{align}
	where \eqref{eq_IL_1}, \eqref{eq_IL_3} follow from the non-negativity of mutual information,\eqref{eq_IL_444}, follows from the independence of $W_{k:K}$, and $Q_n^{[k]}$, \eqref{eq_IL_2} follows from the privacy constraint, \eqref{eq_IL_44}, \eqref{eq_IL_4} follow from the fact that the answer string $A_n^{[k]}$ is a deterministic function of $(Q_n^{[k]},W_{1:K})$, \eqref{eq_IL_5} follows from the chain rule, \eqref{eq_IL_6} follows from the statistical independence of $(\mathbb{H},Q_{1:N}^{[k]}, W_{k:K})$, \eqref{eq_IL_7} is consequence of the decodability of $W_k$ from $( Q_{1:N}^{[k]}, A_{1:N}^{[k]})$, and \eqref{eq_IL_8} is due to the uncoded assumption of the cached bits.
	
	For the second term in \eqref{eq_L2_a}, we have
	\begin{align} 
	I(W_{k:K};Z|W_{1:k-1})&= H\left(Z|W_{1:k-1}\right)-H(Z|W_{1:K}) \\
	&=\left( K-k+1 \right) L r  \label{eq_L2_c}
	\end{align}	
	where \eqref{eq_L2_c} follows from the uncoded nature of the cached bits. 
	
	Combining \eqref{eq_L2_a}, \eqref{eq_L2_b}, and \eqref{eq_L2_c} yields \eqref{eq_L2}. 
\end{Proof}

Now we are ready to derive the general inner bound for arbitrary $K$, $N$, $r$. To obtain this bound, we use Lemma~\ref{lemma_converse1} to find $K$ lower bounds on the length of the undesired portion of the answer strings $D(r)-L(1-r)$. Each lower bound is obtained by varying the index $k$ in the lemma from $k=2$ to $k=K$. Next, we inductively lower bound each result of Lemma~\ref{lemma_converse1} by using Lemma~\ref{lemma_converse2}, precisely $(K-k+1)$ times, to get $K$ explicit lower bounds. This is stated in the following lemma. 
\begin{lemma}\label{Lemma_summary}
	For $N$ and $K$, we have
	\begin{align}
	D(r) \geq L(1-r) \sum_{j=0}^{K+1-k} \frac{1}{N^j} - Lr \sum_{j=0}^{K-k} \frac{K+1-k-j}{N^j}-o(L),
	\end{align}
	where $k=2, \dots, K+1$. 
\end{lemma}
\begin{Proof}
	We have
	\begin{align}
	&D(r)+ o(L)  \notag \\
	&\label{eq_induction1}\qquad \stackrel{\eqref{eq_L1}}{\geq} L(1-r)+ I\left(W_{k:K}; \mathbb{H}, Q_{1:N}^{[k-1]}, A_{1:N}^{[k-1]}|W_{1:k-1}, Z \right) \\
	&\qquad \stackrel{\eqref{eq_L2}}{\geq} L(1-r)+ \frac{L(1-r) - o(L)}{N}  -(K-k+1)Lr   \notag \\
	&\qquad \qquad + \frac{1}{N}  I\left(W_{k+1:K};\mathbb{H}, Q_{1:N}^{[k]}, A_{1:N}^{[k]}|W_{1:k},  Z  \right)   \\
	&\qquad \stackrel{\eqref{eq_L2}}{\geq} L(1-r)\left[1+\frac{1}{N}+\frac{1}{N^2} +o(L)\right] -Lr \left[(K-k+1) +\frac{(K-k)}{N}  \right] \notag \\
	&\qquad \qquad +\frac{1}{N^2} I\left(W_{k+2:K};\mathbb{H}, Q_{1:N}^{[k+1]}, A_{1:N}^{[k+1]}|W_{1:k+1},  Z  \right)  \notag \\
	&\qquad \stackrel{\eqref{eq_L2}}{\geq} \dots \\
	&\qquad \stackrel{\eqref{eq_L2}}{\geq} L(1-r) \sum_{j=0}^{K+1-k} \frac{1}{N^j} - Lr \sum_{j=0}^{K-k} \frac{K+1-k-j}{N^j} + o(L),
	\end{align}
	where \eqref{eq_induction1} follows from Lemma~\ref{lemma_converse1} starting from general index $k$, and the remaining bounding steps correspond to successive application of Lemma~\ref{lemma_converse2}.
\end{Proof}

We conclude the converse proof by dividing by $L$ and taking the limit as $L \rightarrow \infty$, then for $k=2, \cdots, K+1$, we have
\begin{align}\label{beforelast}
D^*(r) \geq (1-r) \sum_{j=0}^{K+1-k} \frac{1}{N^j} - r \sum_{j=0}^{K-k} \frac{K+1-k-j}{N^j} 
\end{align}

Finally, \eqref{beforelast} gives $K$ intersecting line segments, therefore, the normalized download cost is lower bounded by their maximum value  
\begin{align}
D^*(r) \geq \max_{i \in \{2, \cdots, K+1\}} (1-r) \sum_{j=0}^{K+1-i} \frac{1}{N^j} - r \sum_{j=0}^{K-i} \frac{K+1-k-j}{N^j} 
\end{align}

\section{Further Examples}
\subsection{$K=4$ Messages, $N=2$ Databases}

For $K=4$ and $N=2$, we show the achievable PIR schemes for caching ratios $r_1=\frac{1}{15}$ in Table~\ref{N_2_K_4_r=1/15}, $r_2=\frac{1}{5}$ in Table~\ref{N_2_K_4_r=1/5}, and $r_3=\frac{1}{3}$ in Table~\ref{N_2_K_4_r=1/3}. The achievable normalized download costs for these caching ratios are $\frac{22}{15}$, $1$ and $\frac{2}{3}$, respectively. We show the normalized download cost and caching ratio trade off curve in Figure~\ref{K=4,N=2 case}. 

\begin{table}[H] 
	\centering
	\caption{Query table for $K=4$, $N=2$ and $r_1=\frac{1}{15}$}
	\label{N_2_K_4_r=1/15}
	\begin{tabular}{ccc}
		\hline
		\multicolumn{1}{|c|}{$s$}                    & \multicolumn{1}{c|}{DB1}           & \multicolumn{1}{c|}{DB2}           \\ \hline
		\multicolumn{1}{|c|}{\multirow{3}{*}{\rotatebox[origin=c]{90}{\parbox[c]{1cm}{\centering $s=1$}}}} & \multicolumn{1}{c|}{$a_2+b_1$}     & \multicolumn{1}{c|}{$a_5+b_1$}     \\ \cline{2-3} 
		\multicolumn{1}{|c|}{}                       & \multicolumn{1}{c|}{$a_3+c_1$}     & \multicolumn{1}{c|}{$a_6+c_1$}     \\ \cline{2-3} 
		\multicolumn{1}{|c|}{}                       & \multicolumn{1}{c|}{$a_4+d_1$}     & \multicolumn{1}{c|}{$a_7+d_1$}     \\ \hline
		\multicolumn{1}{|c|}{\multirow{2}{*}{}}      & \multicolumn{1}{c|}{$b_2+c_2$}     & \multicolumn{1}{c|}{$b_4+c_4$}     \\ \cline{2-3} 
		\multicolumn{1}{|c|}{}                       & \multicolumn{1}{c|}{$b_3+d_2$}     & \multicolumn{1}{c|}{$b_5+d_4$}      \\ \cline{2-3} 
		\multicolumn{1}{|c|}{}                       & \multicolumn{1}{c|}{$c_3+d_3$}     & \multicolumn{1}{c|}{$c_5+d_5$}      \\ \cline{2-3} 
		\multicolumn{1}{|c|}{}                       & \multicolumn{1}{c|}{$a_8+b_4+c_4$} & \multicolumn{1}{c|}{$a_{11}+b_2+c_2$} \\ \cline{2-3} 
		\multicolumn{1}{|c|}{}                       & \multicolumn{1}{c|}{$a_9+b_5+d_4$} & \multicolumn{1}{c|}{$a_{12}+b_3+d_2$} \\ \cline{2-3} 
		\multicolumn{1}{|c|}{}                       & \multicolumn{1}{c|}{$a_{10}+c_5+d_5$}&\multicolumn{1}{c|}{$a_{13}+c_3+d_3$} \\ \cline{2-3} 
		\multicolumn{1}{|c|}{}                       & \multicolumn{1}{c|}{$b_6+c_6+d_6$} & \multicolumn{1}{c|}{$b_7+c_7+d_7$} \\ \cline{2-3}
		\multicolumn{1}{|c|}{}             & \multicolumn{1}{c|}{$a_{14}+b_7+c_7+d_7$} & \multicolumn{1}{c|}{$a_{15}+b_6+c_6+d_6$} \\ \hline
		&                                    &                                    \\ \cline{2-3} 
		\multicolumn{1}{c|}{}                        & \multicolumn{2}{c|}{$Z=(a_1,b_1,c_1,d_1)$}                                  \\ \cline{2-3} 
	\end{tabular}
\end{table}

\begin{table}[H]
	\centering
	\caption{Query table for $K=4$, $N=2$, $r_2=\frac{1}{5}$}
	\label{N_2_K_4_r=1/5}
	\begin{tabular}{ccc}
		\hline
		\multicolumn{1}{|c|}{$s$}                    & \multicolumn{1}{c|}{DB1}           & \multicolumn{1}{c|}{DB2}           \\ \hline
		\multicolumn{1}{|c|}{\multirow{3}{*}{\rotatebox[origin=c]{90}{\parbox[c]{1cm}{\centering $s=2$}}}} 
		& \multicolumn{1}{c|}{$a_3+b_1+c_1$} & \multicolumn{1}{c|}{$a_6+b_1+c_1$}  \\ \cline{2-3} 
		\multicolumn{1}{|c|}{}                       & \multicolumn{1}{c|}{$a_4+d_1+b_2$} & \multicolumn{1}{c|}{$a_7+d_1+b_2$}   \\ \cline{2-3} 
		\multicolumn{1}{|c|}{}                       & \multicolumn{1}{c|}{$a_5+c_2+d_2$} & \multicolumn{1}{c|}{$a_8+c_2+d_2$}   \\ \hline
		\multicolumn{1}{|c|}{\multirow{2}{*}{}}      & \multicolumn{1}{c|}{$b_3+c_3+d_3$} & \multicolumn{1}{c|}{$b_4+c_4+d_4$}   \\ \cline{2-3} 
		\multicolumn{1}{|c|}{}             & \multicolumn{1}{c|}{$a_9+b_4+c_4+d_4$} & \multicolumn{1}{c|}{$a_{10}+b_3+c_3+d_3$} \\ \hline 
		&                                    &                                    \\ \cline{2-3} 
		\multicolumn{1}{c|}{}                        & \multicolumn{2}{c|}{$Z=(a_1,a_2,b_1,b_2,c_1,c_2,d_1,d_2)$}                                  \\ \cline{2-3} 
	\end{tabular}
\end{table}

\begin{table}[H]
	\centering
	\caption{Query table for $K=4$, $N=2$, $r_3=\frac{1}{3}$}
	\label{N_2_K_4_r=1/3}
	\begin{tabular}{ccc}
		\hline
		\multicolumn{1}{|c|}{$s$}   & \multicolumn{1}{c|}{DB1}              & \multicolumn{1}{c|}{DB2}           \\ \hline
		\multicolumn{1}{|c|}{$s=3$} & \multicolumn{1}{c|}{$a_2+b_1+c_1+d_1$} & \multicolumn{1}{c|}{$a_3+b_1+c_1+d_1$} \\ \hline 
		&                                       &                                    \\ \cline{2-3} 
		\multicolumn{1}{c|}{}       & \multicolumn{2}{c|}{$Z=(a_1,b_1,c_1,d_1)$}                                     \\ \cline{2-3} 
	\end{tabular}
\end{table}

\subsection{$K=4$ Messages, $N=3$ Databases}

\begin{figure}[t]
	\centering
	\epsfig{file=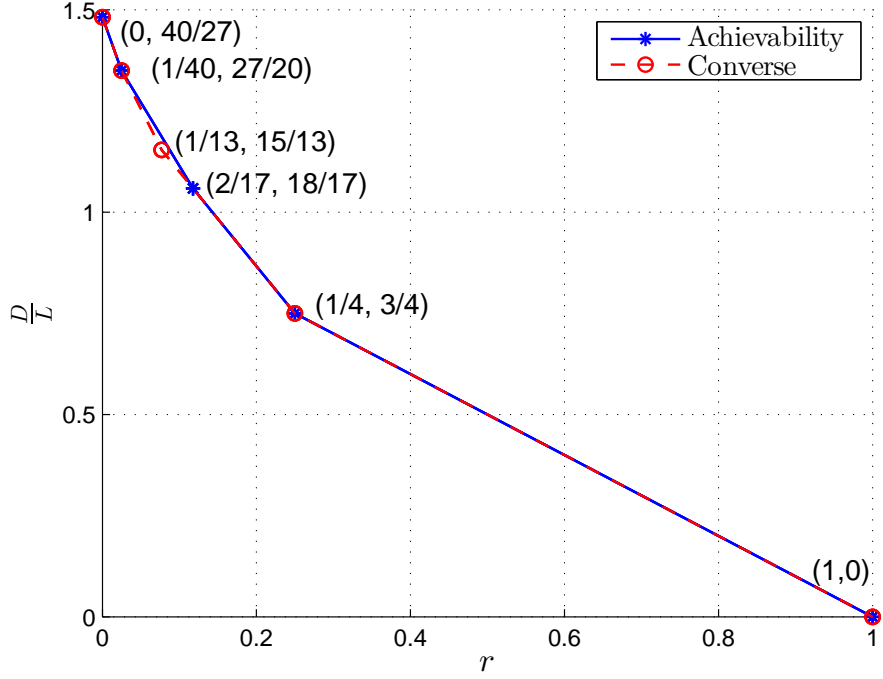,width=0.8\textwidth}
	\caption{Inner and outer bounds for $K=4$, $N=3$.}
	\label{K=3_4_N=2 case}
\end{figure}

For $K=4$ and $N=3$, we show the achievable PIR schemes for caching ratios $r_1=\frac{1}{40}$ in Table~\ref{N_3_K_4_r=1/40}, $r_2=\frac{2}{17}$ in Table~\ref{N_3_K_4_r=2/17}, and $r_3=\frac{1}{4}$ in Table~\ref{N_3_K_4_r=1/4}. We show the normalized download cost and caching ratio trade off in Figure~\ref{K=3_4_N=2 case}. The achievable normalized download costs for these caching ratios are $\frac{27}{20}$, $\frac{18}{17}$ and $\frac{3}{4}$, respectively. By comparing Figure~\ref{K=3_4_N=2 case} with Figure~\ref{K=4,N=2 case}, we observe that, for fixed $K$, as $N$ grows, the gap between the achievable bound and the converse bound shrinks. This observation will be specified in Section~\ref{Sec_gap}. 

\begin{table}[H]
	\centering
	\caption{Query table for $K=4$, $N=3$, $r_1=\frac{1}{40}$}
	\label{N_3_K_4_r=1/40}
	\begin{tabular}{cccc}
		\hline
		\multicolumn{1}{|c|}{$s$}        & \multicolumn{1}{c|}{DB1}     & \multicolumn{1}{c|}{DB2}    &\multicolumn{1}{c|}{DB3}       \\ \hline
		\multicolumn{1}{|c|}{\multirow{3}{*}{\rotatebox[origin=c]{90}{\parbox[c]{1cm}{\centering $s=1$}}}} 
		&\multicolumn{1}{c|}{$a_2+b_1$} &\multicolumn{1}{c|}{$a_5+b_1$} &\multicolumn{1}{c|}{$a_8+b_1$}  \\\cline{2-4}  
		\multicolumn{1}{|c|}{}   &\multicolumn{1}{c|}{$a_3+c_1$} &\multicolumn{1}{c|}{$a_6+c_1$} &\multicolumn{1}{c|}{$a_9+c_1$}   \\\cline{2-4} 
		\multicolumn{1}{|c|}{}   &\multicolumn{1}{c|}{$a_4+d_1$} &\multicolumn{1}{c|}{$a_7+d_1$} &\multicolumn{1}{c|}{$a_{10}+d_1$}\\\hline
		\multicolumn{1}{|c|}{}  & \multicolumn{1}{c|}{$b_2+c_2$} &\multicolumn{1}{c|}{$b_4+c_4$} &\multicolumn{1}{c|}{$b_6+c_6$}   \\\cline{2-4}
		\multicolumn{1}{|c|}{}  & \multicolumn{1}{c|}{$b_3+d_2$} &\multicolumn{1}{c|}{$b_5+d_4$} &\multicolumn{1}{c|}{$b_7+d_6$}   \\\cline{2-4}
		\multicolumn{1}{|c|}{}  & \multicolumn{1}{c|}{$c_3+d_3$} &\multicolumn{1}{c|}{$c_5+d_5$} &\multicolumn{1}{c|}{$c_7+d_7$}   \\\cline{2-4}
		\multicolumn{1}{|c|}{} 
		&\multicolumn{1}{c|}{$a_{11}+b_4+d_4$} & \multicolumn{1}{c|}{$a_{17}+b_2+c_2$} & \multicolumn{1}{c|}{$a_{23}+b_2+c_2$} \\\cline{2-4} 
		\multicolumn{1}{|c|}{} 
		&\multicolumn{1}{c|}{$a_{12}+b_5+d_4$} & \multicolumn{1}{c|}{$a_{18}+b_3+d_2$} & \multicolumn{1}{c|}{$a_{24}+b_3+d_2$} \\\cline{2-4}   
		\multicolumn{1}{|c|}{} 
		&\multicolumn{1}{c|}{$a_{13}+c_5+d_5$} & \multicolumn{1}{c|}{$a_{19}+c_3+d_3$} & \multicolumn{1}{c|}{$a_{25}+c_3+d_3$} \\\cline{2-4}    
		\multicolumn{1}{|c|}{} 
		&\multicolumn{1}{c|}{$a_{14}+b_6+c_6$} & \multicolumn{1}{c|}{$a_{20}+b_6+c_6$} & \multicolumn{1}{c|}{$a_{26}+b_4+c_4$}\\\cline{2-4}      
		\multicolumn{1}{|c|}{} 
		&\multicolumn{1}{c|}{$a_{15}+b_7+d_6$} & \multicolumn{1}{c|}{$a_{21}+b_7+d_6$} & \multicolumn{1}{c|}{$a_{27}+b_5+d_4$}\\\cline{2-4}      
		\multicolumn{1}{|c|}{}
		&\multicolumn{1}{c|}{$a_{16}+c_7+d_7$} & \multicolumn{1}{c|}{$a_{22}+c_7+d_7$} & \multicolumn{1}{c|}{$a_{28}+c_5+d_5$}\\\cline{2-4}       
		\multicolumn{1}{|c|}{}
		&\multicolumn{1}{c|}{$b_8+c_8+d_8$}&\multicolumn{1}{c|}{$b_{10}+c_{10}+d_{10}$}&\multicolumn{1}{c|}{$b_{12}+c_{12}+d_{12}$}\\\cline{2-4}  \multicolumn{1}{|c|}{}
		&\multicolumn{1}{c|}{$b_9+c_9+d_9$}&\multicolumn{1}{c|}{$b_{11}+c_{11}+d_{11}$}&\multicolumn{1}{c|}{$b_{13}+c_{13}+d_{13}$}\\\cline{2-4}  
		\multicolumn{1}{|c|}{}
		&\multicolumn{1}{c|}{$a_{29}+b_{10}+c_{10}+d_{10}$}&\multicolumn{1}{c|}{$a_{33}+b_{8}+c_{8}+d_{8}$}
		&\multicolumn{1}{c|}{$a_{37}+b_{8}+c_{8}+d_{8}$}\\\cline{2-4}  
		\multicolumn{1}{|c|}{}
		&\multicolumn{1}{c|}{$a_{30}+b_{11}+c_{11}+d_{11}$}&\multicolumn{1}{c|}{$a_{34}+b_{9}+c_{9}+d_{9}$}
		&\multicolumn{1}{c|}{$a_{38}+b_{9}+c_{9}+d_{9}$}\\\cline{2-4}
		\multicolumn{1}{|c|}{}
		&\multicolumn{1}{c|}{$a_{31}+b_{12}+c_{12}+d_{12}$}&\multicolumn{1}{c|}{$a_{35}+b_{12}+c_{12}+d_{12}$}
		&\multicolumn{1}{c|}{$a_{39}+b_{10}+c_{10}+d_{10}$}\\\cline{2-4}
		\multicolumn{1}{|c|}{}
		&\multicolumn{1}{c|}{$a_{32}+b_{13}+c_{13}+d_{13}$}&\multicolumn{1}{c|}{$a_{36}+b_{13}+c_{13}+d_{13}$}
		&\multicolumn{1}{c|}{$a_{40}+b_{11}+c_{11}+d_{11}$}\\\hline
		&                                                                                \\ \cline{2-4} 
		\multicolumn{1}{c|}{}   & \multicolumn{3}{c|}{$Z=(a_1,b_1,c_1,d_1)$}                 \\ \cline{2-4} 
	\end{tabular}
\end{table}

\begin{table}[H]
	\centering
	\caption{Query table for $K=4$, $N=3$, $r_2=\frac{2}{17}$}
	\label{N_3_K_4_r=2/17}
	\begin{tabular}{cccc}
		\hline
		\multicolumn{1}{|c|}{$s$}        & \multicolumn{1}{c|}{DB1}     & \multicolumn{1}{c|}{DB2}    &\multicolumn{1}{c|}{DB3}       \\ \hline
		\multicolumn{1}{|c|}{\multirow{3}{*}{\rotatebox[origin=c]{90}{\parbox[c]{1cm}{\centering $s=2$}}}} 
		&\multicolumn{1}{c|}{$a_3+b_1+c_1$} &\multicolumn{1}{c|}{$a_6+b_1+c_1$} &\multicolumn{1}{c|}{$a_9+b_1+c_1$}  \\\cline{2-4}  
		\multicolumn{1}{|c|}{}   
		&\multicolumn{1}{c|}{$a_4+d_1+b_2$} &\multicolumn{1}{c|}{$a_7+d_1+b_2$} &\multicolumn{1}{c|}{$a_{10}+d_1+b_2$}  \\\cline{2-4}  
		\multicolumn{1}{|c|}{}   
		&\multicolumn{1}{c|}{$a_5+c_2+d_2$} &\multicolumn{1}{c|}{$a_8+c_2+d_2$} &\multicolumn{1}{c|}{$a_{11}+c_2+d_2$}  \\\hline  
		\multicolumn{1}{|c|}{}   
		&\multicolumn{1}{c|}{$b_3+c_3+d_3$} &\multicolumn{1}{c|}{$b_4+c_4+d_4$} &\multicolumn{1}{c|}{$b_5+c_5+d_5$}  \\\cline{2-4}
		\multicolumn{1}{|c|}{}   
		&\multicolumn{1}{c|}{$a_{12}+b_{4}+c_{4}+d_{4}$}&\multicolumn{1}{c|}{$a_{14}+b_{3}+c_{3}+d_{3}$}
		&\multicolumn{1}{c|}{$a_{16}+b_{3}+c_{3}+d_{3}$}\\\cline{2-4} 
		\multicolumn{1}{|c|}{}   
		&\multicolumn{1}{c|}{$a_{13}+b_{5}+c_{5}+d_{5}$}&\multicolumn{1}{c|}{$a_{15}+b_{5}+c_{5}+d_{5}$}
		&\multicolumn{1}{c|}{$a_{17}+b_{4}+c_{4}+d_{4}$}\\\hline 
		&                                                                                \\ \cline{2-4} 
		\multicolumn{1}{c|}{}   & \multicolumn{3}{c|}{$Z=(a_1,a_2,b_1,b_2,c_1,c_2,d_1,d_2)$}                 \\ \cline{2-4} 
	\end{tabular}
\end{table}

\begin{table}[H]
	\centering
	\caption{Query table for $K=4$, $N=3$, $r_3=\frac{1}{4}$}
	\label{N_3_K_4_r=1/4}
	\begin{tabular}{cccc}
		\hline
		\multicolumn{1}{|c|}{$s$}   & \multicolumn{1}{c|}{DB1}          & \multicolumn{1}{c|}{DB2}   & \multicolumn{1}{c|}{DB3}     \\ \hline
		\multicolumn{1}{|c|}{$s=3$} 
		& \multicolumn{1}{c|}{$a_2+b_1+c_1+d_1$} & \multicolumn{1}{c|}{$a_3+b_1+c_1+d_1$}  & \multicolumn{1}{c|}{$a_4+b_1+c_1+d_1$} \\ \hline
		&                                       &                                    \\ \cline{2-4} 
		\multicolumn{1}{c|}{}     & \multicolumn{3}{c|}{$Z=(a_1,b_1,c_1,d_1)$}                                     \\ \cline{2-4} 
	\end{tabular}
\end{table}

\subsection{$K=5$, $K=10$ and $K=100$ Messages, $N=2$ Databases}

\begin{figure}[t]
\centering
\epsfig{file=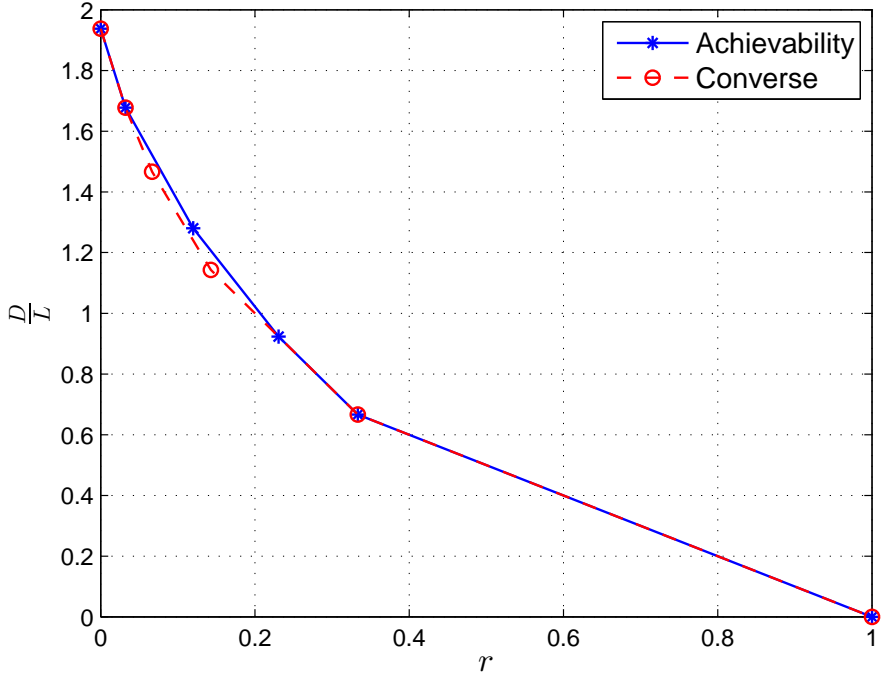,width=0.8\textwidth}
\caption{Inner and outer bounds for $K=5$, $N=2$.}
\label{fig_N2K5}
\end{figure}

\begin{figure}[t]
	\centering
	\epsfig{file=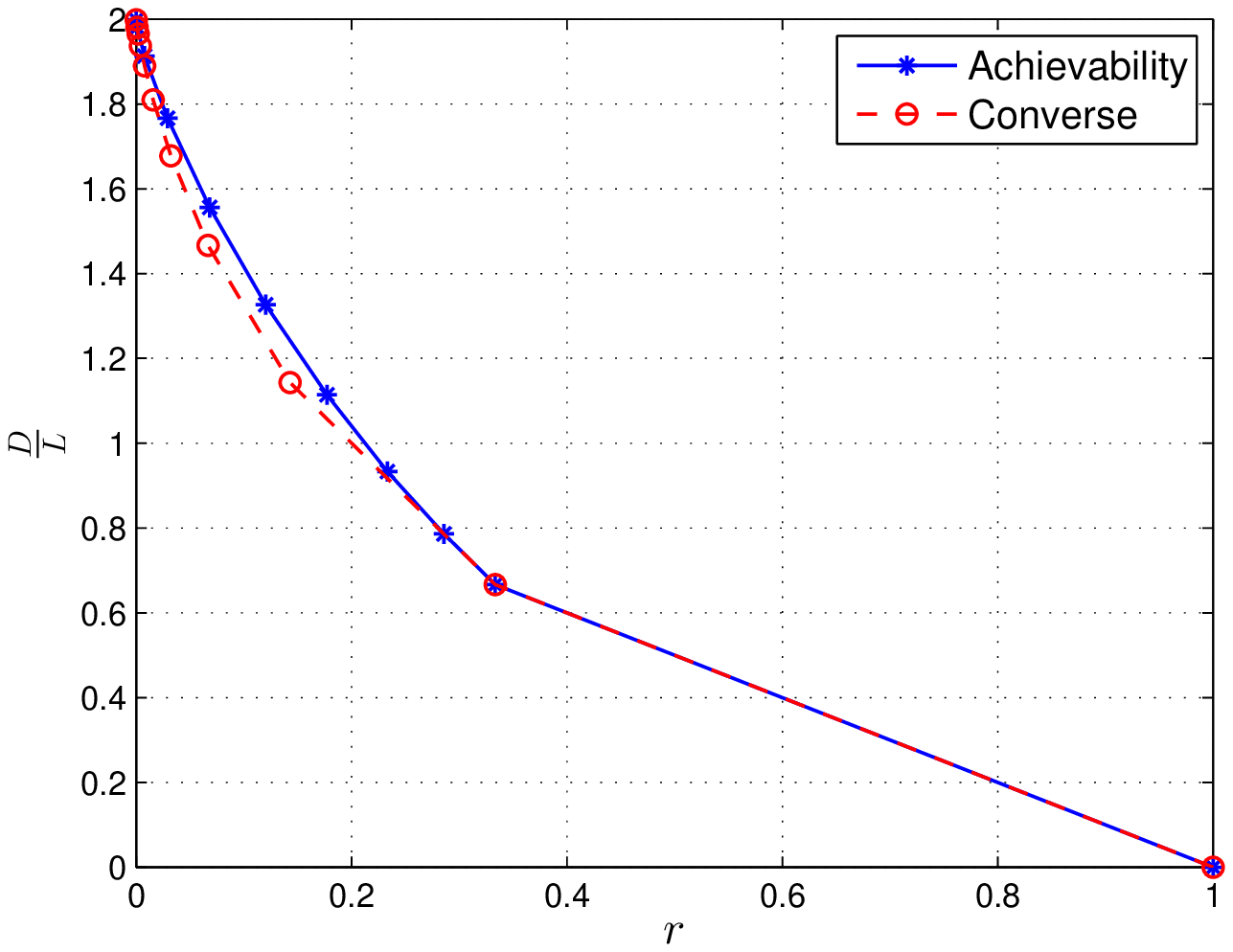,width=0.8\textwidth}
	\caption{Inner and outer bounds for $K=10$, $N=2$.}
	\label{fig_N2K10}
\end{figure}

\begin{figure}[t]
	\centering
	\epsfig{file=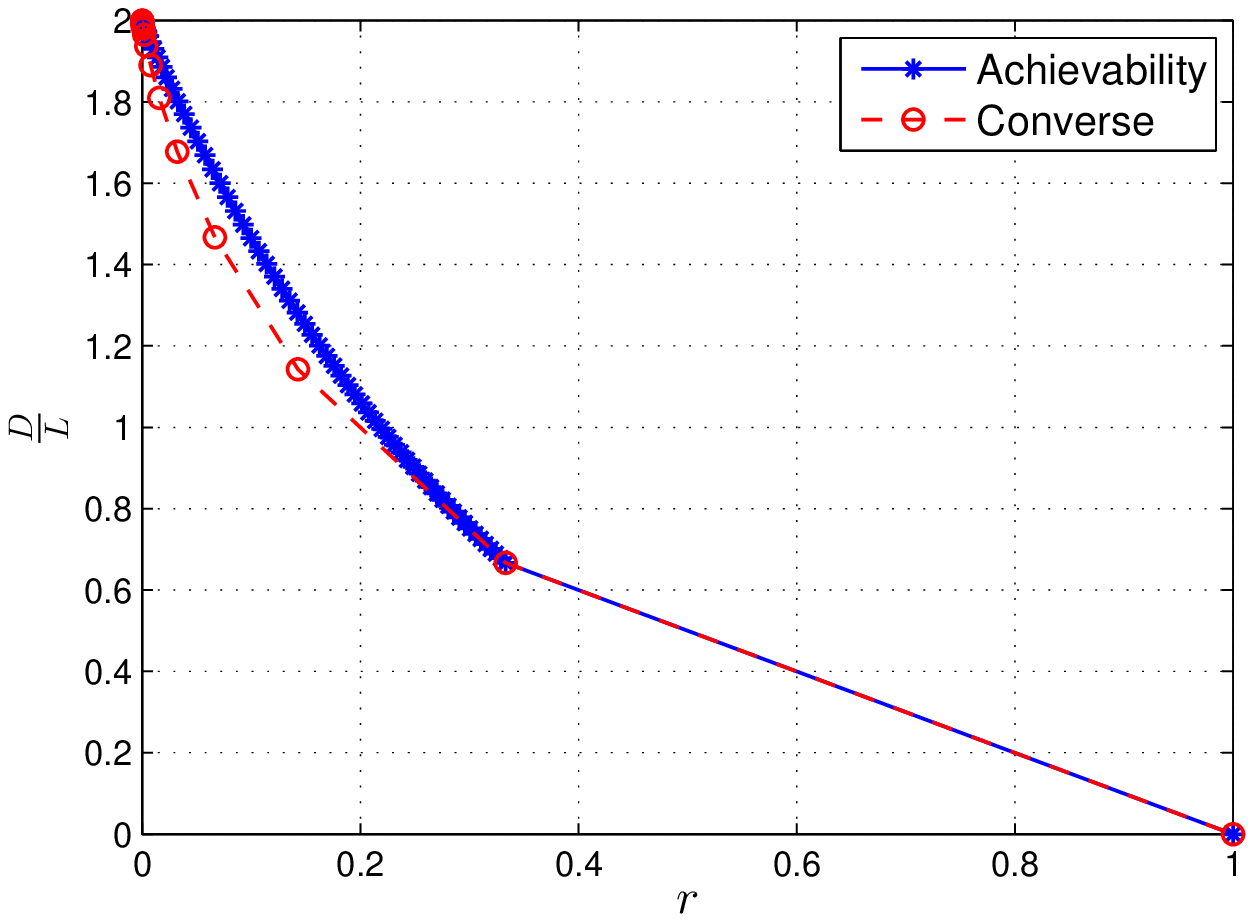,width=0.8\textwidth}
	\caption{Inner and outer bounds for $K=100$, $N=2$.}
		\label{fig_N2K100}
\end{figure}

For $N=2$, we show the numerical results for the inner and outer bounds for $K=5$, $K=10$ and $K=100$  in Figures~\ref{fig_N2K5}, \ref{fig_N2K10} and \ref{fig_N2K100}. For fixed $N$ as $K$ grows, the gap between the achievable bound and converse bound increases. This observation will be specified in Section~\ref{Sec_gap}. 

\subsection{$K=5$, $K=10$ and $K=100$ Messages, $N=3$ Databases}

\begin{figure}[t]
	\centering
	\epsfig{file=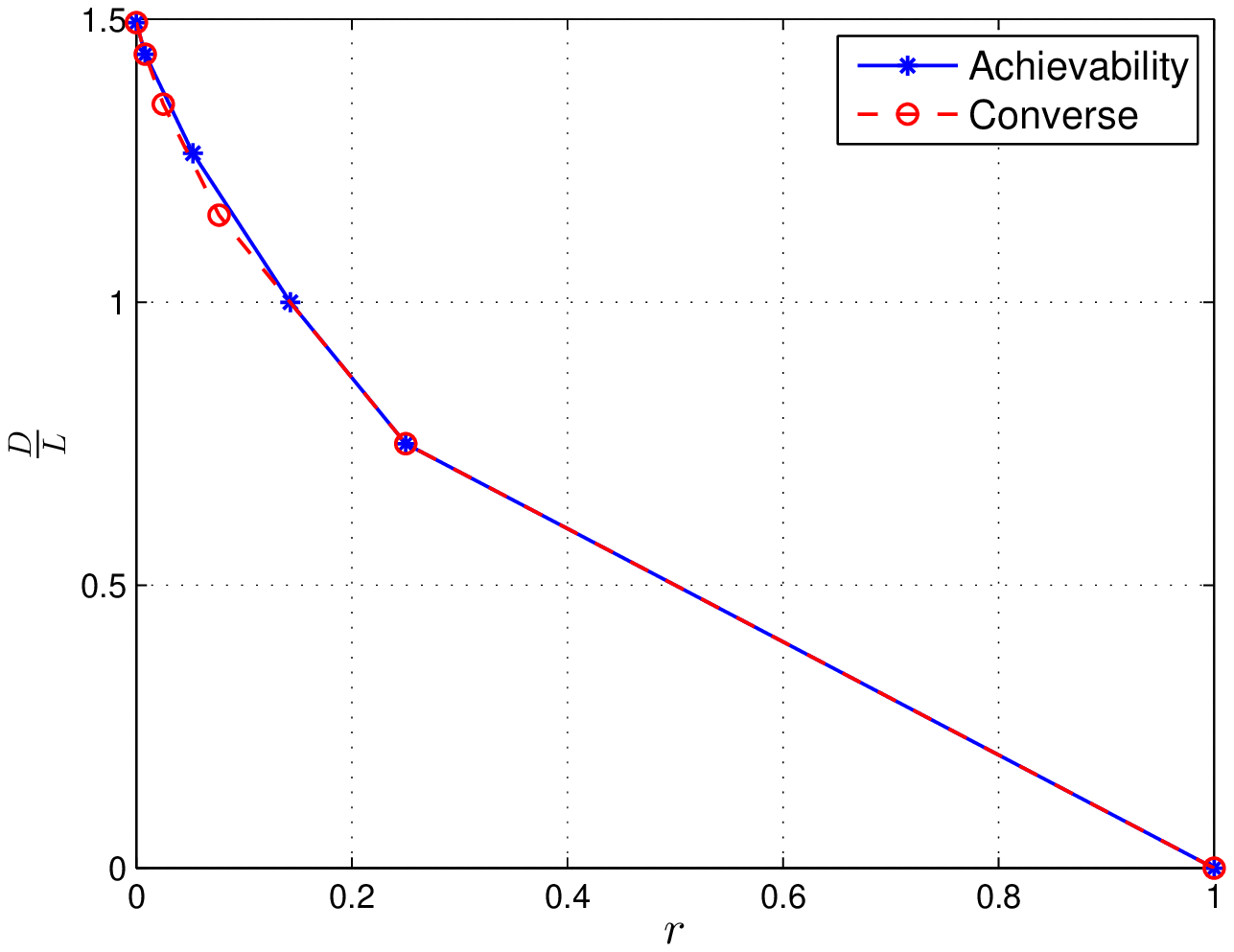,width=0.8\textwidth}
	\caption{Inner and outer bounds for $K=5$, $N=3$.}
	\label{fig_N3K5}
\end{figure}

\begin{figure}[t]
	\centering
	\epsfig{file=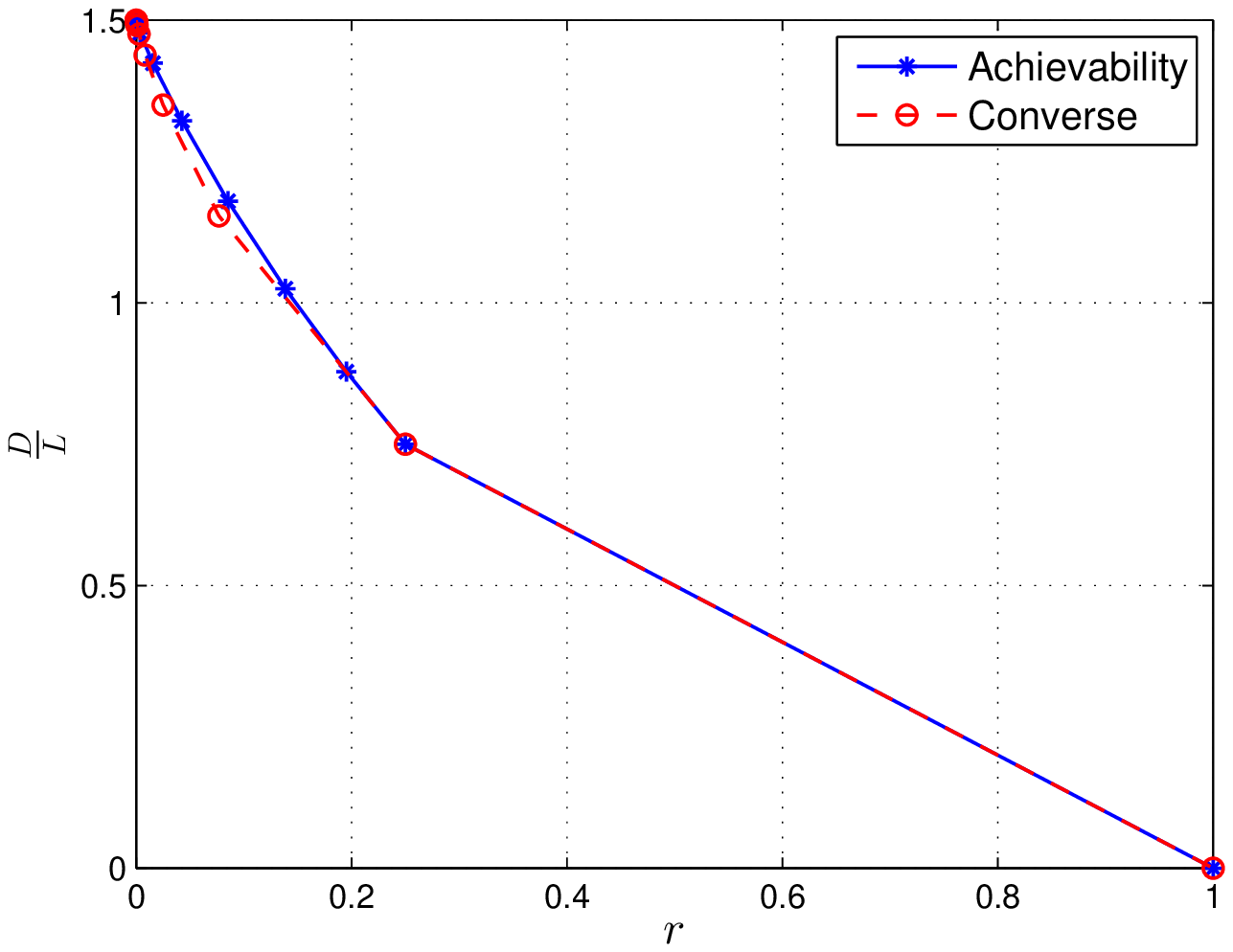,width=0.8\textwidth}
	\caption{Inner and outer bounds for $K=10$, $N=3$.}
	\label{fig_N3K10}
\end{figure}

\begin{figure}[t]
	\centering
	\epsfig{file=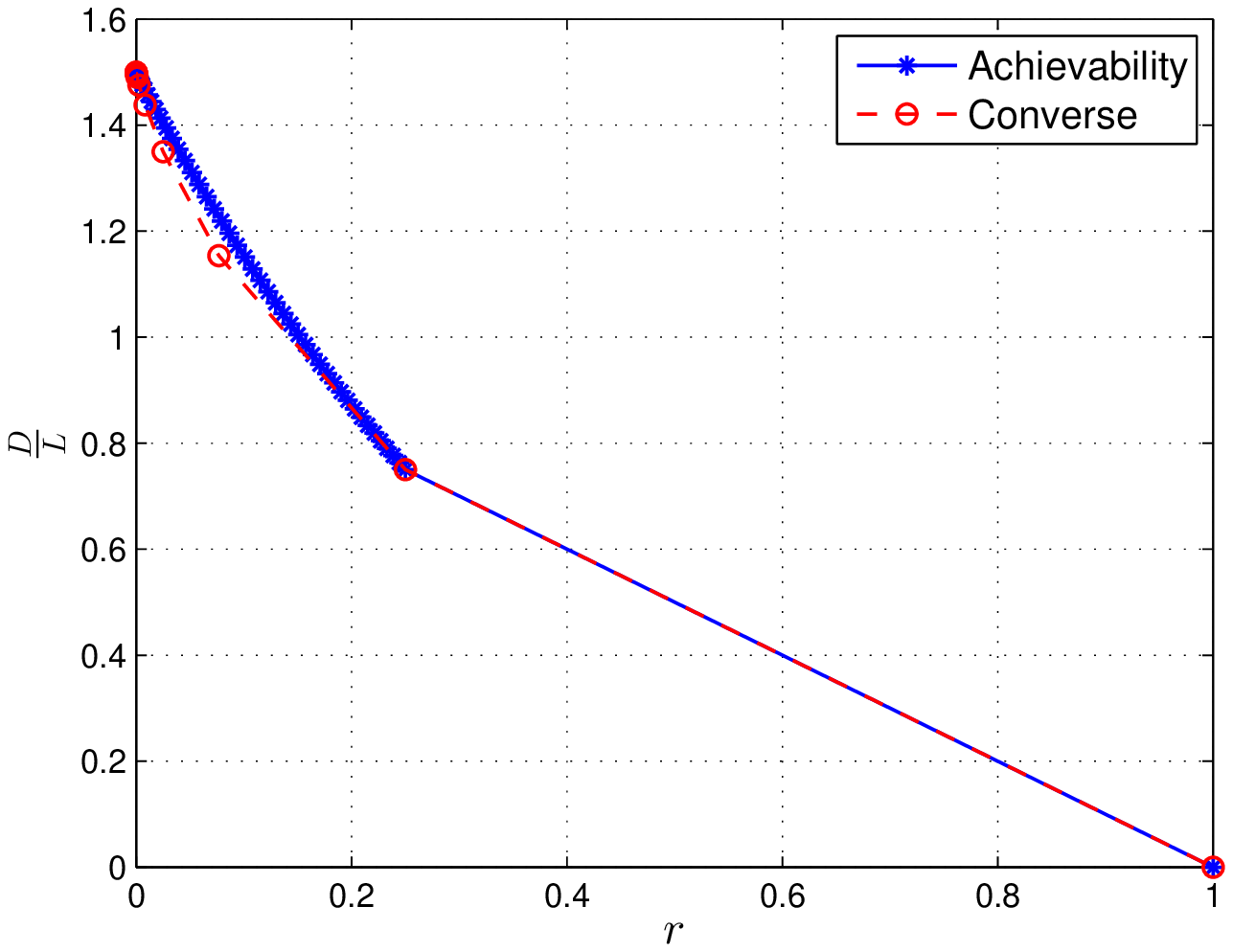,width=0.8\textwidth}
	\caption{Inner and outer bounds for $K=100$, $N=3$.}
	\label{fig_N3K100}
\end{figure}

For $N=3$, we show the numerical results for the inner and outer bounds for $K=5$, $K=10$ and $K=100$  in Figures~\ref{fig_N3K5}, \ref{fig_N3K10} and \ref{fig_N3K100}. For fixed $N$ as $K$ grows, the gap between the achievable bound and converse bound increases. This observation will be specified in Section~\ref{Sec_gap}. 

\section{Gap Analysis} \label{Sec_gap}

\begin{figure}[t]
	\centering
	\epsfig{file=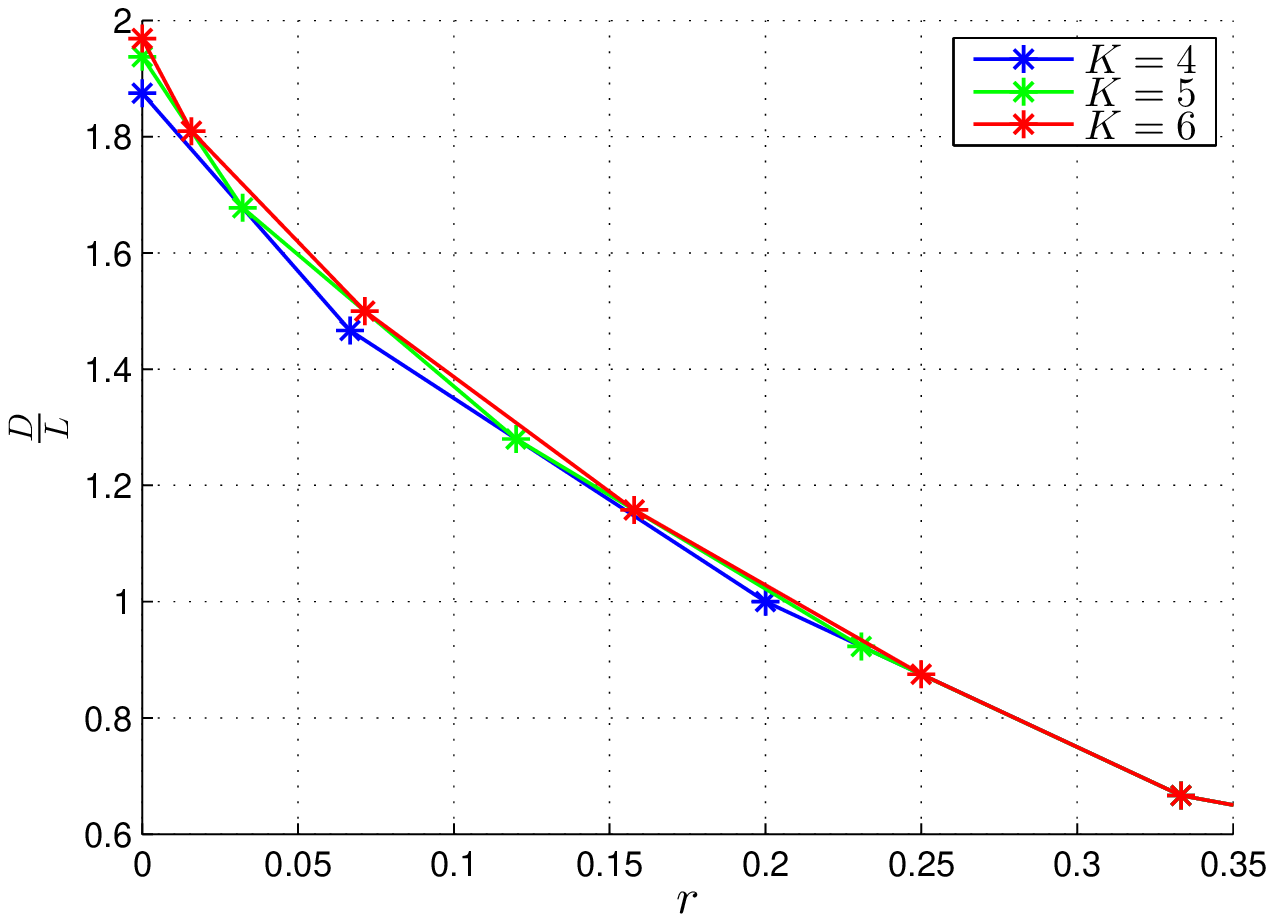,width=0.8\textwidth}
	\caption{Outer bounds for $N=2$, $K=4$, $K=5$ and $K=6$.}
	\label{fig_N2K456}
\end{figure}

In this section, we analyze the gap between the achievability and converse bounds for general $N$, $K$, and $r$, and show that the worst-case gap, which happens when $N=2$ and $K\rightarrow \infty$, is at most $\frac{1}{6}$. We start this section with an interesting property for the monotonicity of the achievable bounds. We first see an example. For $N=2$, $K=4$, $K=5$ and $K=6$, the achievable bounds are shown in Figure~\ref{fig_N2K456}. The achievable bound for $K=6$ is above the achievable bound for $K=5$, and the achievable bound for $K=5$ is above the achievable bound for $K=4$. By denoting $r_s^{(K)}$ as the caching ratio with total $K$ messages and parameter $s$ (see \eqref{r_exp}), we observe that $(r^{(5)}_1, \bar{D}(r^{(5)}_1) )$ falls on the line connecting  $(r^{(4)}_0, \bar{D}(r^{(4)}_0) )$  and $(r^{(4)}_1, \bar{D}(r^{(4)}_1) )$. This observation is general, $(r_s^{(K+1)},\bar{D}(r_s^{(K+1)}) )$ falls on the line connecting $(r_{s-1}^{(K)},\bar{D}(r_{s-1}^{(K)}))$ and $(r_{s}^{(K)},\bar{D}(r_{s}^{(K)}))$. We state and prove this observation in the following lemma. 

\begin{lemma}[Monotonicity of the achievable bounds]
In cache-aided PIR with uncoded and unknown prefetching, for fixed number of databases $N$, if the number of messages $K$ increases, then the achievable normalized download cost increases. Furthermore, we have 
\begin{align}
r_s^{(K+1)} &= \alpha r_{s-1}^{(K)} + (1-\alpha) r_s^{(K)}, \label{m1} \\
\bar{D}(r_s^{(K+1)}) &=  \alpha \bar{D}(r_{s-1}^{(K)}) + (1-\alpha) \bar{D}(r_s^{(K)}),  \label{m2}
\end{align}
where $0\leq \alpha \leq 1$. 
\end{lemma}
\begin{Proof}
To show \eqref{m2} is equivalent to show
\begin{align} \label{m2_1}
\bar{D}(r_s^{(K+1)}) - \bar{D}(r_s^{(K)}) = \alpha\left( \bar{D}(r_{s-1}^{(K)}) - \bar{D}(r_s^{(K)}) \right). 
\end{align}
From \eqref{m1}, we have 
\begin{align} \label{alpha}
\alpha= \frac{r_s^{(K)}-r_s^{(K+1)} }{r_{s}^{(K)}-r_{s-1}^{(K)}}. 
\end{align}
Therefore, to show \eqref{m2_1} is equivalent to show 
\begin{align} \label{m2_2}
\left(r_s^{(K)} - r_{s-1}^{(K)} \right) \left( \bar{D}(r_{s}^{(K+1)}) - \bar{D}(r_s^{(K)}) \right)
=\left(r_s^{(K)} - r_s^{(K+1)} \right) \left( \bar{D}(r_{s-1}^{(K)}) - \bar{D}(r_s^{(K)}) \right). 
\end{align}
Let $\bar{D}(r_s^{(K)}) = \frac{D_s^{(K)}}{L_s^{(K)}}$, where 
\begin{align}
L_s^{(K)}&=\binom{K-2}{s-1}+\sum_{i=0}^{K-1-s} \binom{K-1}{s+i}(N-1)^iN, \label{mm1} \\
D_s^{(K)}&=\sum_{i=0}^{K-1-s} \binom{K}{s+1+i}(N-1)^i N. \label{mm2}
\end{align}
To show \eqref{m2_2} is equivalent to show  
\begin{align}
\left[\frac{\binom{K-2}{s-1}}{L_s^{(K)}} - \frac{\binom{K-2}{s-2}}{L_{s-1}^{(K)}} \right]
\left[  \frac{D_s^{(K+1)}}{L_s^{(K+1)}} - \frac{D_s^{(K)}}{L_s^{(K)}}   \right]
= \left[\frac{\binom{K-2}{s-1}}{L_s^{(K)}} - \frac{\binom{K-1}{s-1}}{L_{s}^{(K+1)}} \right]
\left[  \frac{D_{s-1}^{(K)}}{L_{s-1}^{(K)}} - \frac{D_s^{(K)}}{L_s^{(K)}}   \right],  \label{mm3}
\end{align}
which is obtained by using \eqref{r_exp}, \eqref{eq_outer}, \eqref{mm1} and \eqref{mm2}. Expanding \eqref{mm3}, we have
\begin{align}
&\frac{\binom{K-2}{s-1}}{L_s^{(K)}}  \frac{D_{s}^{(K+1)}}{L_{s}^{(K+1)}} 
- \frac{\binom{K-2}{s-2}}{L_{s-1}^{(K)}} \frac{D_s^{(K+1)}}{L_s^{(K+1)}} 
+ \frac{\binom{K-2}{s-2}}{L_{s-1}^{(K)}}   \frac{D_s^{(K)}}{L_s^{(K)}} 
= \frac{\binom{K-2}{s-1}}{L_s^{(K)}}  \frac{D_{s-1}^{(K)}}{L_{s-1}^{(K)}}
-\frac{\binom{K-1}{s-1}}{L_{s}^{(K+1)}}\frac{D_{s-1}^{(K)}}{L_{s-1}^{(K)}}  
+ \frac{\binom{K-1}{s-1}}{L_{s}^{(K+1)}}  \frac{D_s^{(K)}}{L_s^{(K)}}. \label{mm4} 
\end{align}
Multiplying $L_s^{(K)} L_{s-1}^{(K)} L_{s}^{(K+1)}$ to both side of \eqref{mm4}, we have
\begin{align}
&\binom{K-2}{s-1}D_{s}^{(K+1)}L_{s-1}^{(K)} + \binom{K-1}{s-1}D_{s-1}^{(K)}L_{s}^{(K)} +\binom{K-2}{s-2}D_s^{(K)}L_s^{(K+1)} \notag \\
&\quad = \binom{K-2}{s-1} D_{s-1}^{(K)}L_{s}^{(K+1)}+ \binom{K-2}{s-2} D_s^{(K+1)}L_{s}^{(K)}+\binom{K-1}{s-1}D_s^{(K)}L_{s-1}^{(K)}. 
\end{align}
By using \eqref{mm1} and \eqref{mm2}, we further have
\begin{align}
&\binom{K-2}{s-1} \left[\sum_{i=0}^{K-s} \binom{K+1}{s+1+i}(N-1)^i N  \right] 
\left[\binom{K-2}{s-2}+\sum_{i=0}^{K-s} \binom{K-1}{s-1+i}(N-1)^iN \right] \notag \\ 
&\quad +\binom{K-1}{s-1} \left[\sum_{i=0}^{K-s} \binom{K}{s+i}(N-1)^i N  \right] 
\left[\binom{K-2}{s-1}+\sum_{i=0}^{K-1-s} \binom{K-1}{s+i}(N-1)^iN\right] \notag \\
&\quad +\binom{K-2}{s-2} \left[\sum_{i=0}^{K-1-s} \binom{K}{s+1+i}(N-1)^i N\right] 
\left[\binom{K-1}{s-1}+\sum_{i=0}^{K-s} \binom{K}{s+i}(N-1)^iN \right] \notag \\
&= \binom{K-2}{s-1} \left[ \sum_{i=0}^{K-s} \binom{K}{s+i}(N-1)^i N \right] 
\left[\binom{K-1}{s-1}+\sum_{i=0}^{K-s} \binom{K}{s+i}(N-1)^iN \right] \notag \\
&\quad + \binom{K-2}{s-2} \left[\sum_{i=0}^{K-s} \binom{K+1}{s+1+i}(N-1)^i N \right] 
\left[\binom{K-2}{s-1}+\sum_{i=0}^{K-1-s} \binom{K-1}{s+i}(N-1)^iN \right] \notag \\
&\quad + \binom{K-1}{s-1} \left[\sum_{i=0}^{K-1-s} \binom{K}{s+1+i}(N-1)^i N\right]
\left[\binom{K-2}{s-2}+\sum_{i=0}^{K-s} \binom{K-1}{s-1+i}(N-1)^iN\right]. 
\end{align}
By canceling same terms on both sides, we have 
\begin{align}
&\binom{K-2}{s-1} \left[\sum_{i=0}^{K-s} \binom{K+1}{s+1+i}(N-1)^i   \right] 
\left[\sum_{i=0}^{K-s} \binom{K-1}{s-1+i}(N-1)^i \right] \notag \\ 
&\quad +\binom{K-1}{s-1} \left[\sum_{i=0}^{K-s} \binom{K}{s+i}(N-1)^i   \right] 
\left[\sum_{i=0}^{K-1-s} \binom{K-1}{s+i}(N-1)^i\right] \notag \\
&\quad +\binom{K-2}{s-2} \left[\sum_{i=0}^{K-1-s} \binom{K}{s+1+i}(N-1)^i \right] 
\left[\sum_{i=0}^{K-s} \binom{K}{s+i}(N-1)^i \right] \notag \\
&= \binom{K-2}{s-1} \left[ \sum_{i=0}^{K-s} \binom{K}{s+i}(N-1)^i  \right] 
\left[\sum_{i=0}^{K-s} \binom{K}{s+i}(N-1)^i \right] \notag \\
&\quad + \binom{K-2}{s-2} \left[\sum_{i=0}^{K-s} \binom{K+1}{s+1+i}(N-1)^i  \right] 
\left[\sum_{i=0}^{K-1-s} \binom{K-1}{s+i}(N-1)^i \right] \notag \\
&\quad + \binom{K-1}{s-1} \left[\sum_{i=0}^{K-1-s} \binom{K}{s+1+i}(N-1)^i \right]
\left[\sum_{i=0}^{K-s} \binom{K-1}{s-1+i}(N-1)^i\right].
\end{align}
By using the fact that $\binom{K}{s}=\binom{K-1}{s}+\binom{K-1}{s-1}$, we have
\begin{align}  
&\binom{K-2}{s-1} \left[\sum_{i=0}^{K-s} \left( \binom{K}{s+1+i} + \binom{K}{s+i}  \right)   (N-1)^i   \right]
\left[\sum_{i=0}^{K-s} \binom{K-1}{s-1+i}(N-1)^i \right] \notag \\
&\quad + \left(\binom{K-2}{s-1}+\binom{K-2}{s-2}\right) \left[\sum_{i=0}^{K-s} \binom{K}{s+i}(N-1)^i   \right] 
\left[\sum_{i=0}^{K-1-s} \binom{K-1}{s+i}(N-1)^i\right] \notag \\
&\quad + \binom{K-2}{s-2} \left[\sum_{i=0}^{K-1-s} \binom{K}{s+1+i}(N-1)^i \right] 
\left[\sum_{i=0}^{K-s}  \left(\binom{K-1}{s+i}+\binom{K-1}{s+i-1}  \right)   (N-1)^i \right] \notag \\
&= \binom{K-2}{s-1} \left[ \sum_{i=0}^{K-s} \binom{K}{s+i}(N-1)^i  \right]  
\left[\sum_{i=0}^{K-s}  \left(\binom{K-1}{s+i} +\binom{K-1}{s+i-1}  \right) (N-1)^i \right] \notag \\
&\quad + \binom{K-2}{s-2} \left[\sum_{i=0}^{K-s} \left(\binom{K}{s+1+i}+\binom{K}{s+i} \right) (N-1)^i  \right]
\left[\sum_{i=0}^{K-1-s} \binom{K-1}{s+i}(N-1)^i \right] \notag \\
&\quad + \left(\binom{K-2}{s-1}+\binom{K-2}{s-2}   \right) \left[\sum_{i=0}^{K-1-s} \binom{K}{s+1+i}(N-1)^i \right]
\left[\sum_{i=0}^{K-s} \binom{K-1}{s-1+i}(N-1)^i\right]. \label{m2_3}
\end{align}
Since the left hand side of \eqref{m2_3} is equal to the right hand side of \eqref{m2_3}, \eqref{m2} holds. 

To show $\alpha \geq 0$, since $r_{s}^{(K)}>r_{s-1}^{(K)}$ in \eqref{alpha}, it suffices to show that $r_s^{(K)} \geq r_{s}^{(K+1)}$. From \eqref{r_exp}, it is equivalent to show that 
\begin{align}
\frac{\binom{K-2}{s-1}}{\binom{K-2}{s-1}+\sum_{i=0}^{K-1-s} \binom{K-1}{s+i}(N-1)^iN} &\geq 
 \frac{\binom{K-1}{s-1}}{\binom{K-1}{s-1}+\sum_{i=0}^{K-s} \binom{K}{s+i}(N-1)^iN}.
\end{align}
By using the fact that $\binom{K}{s}=\binom{K-1}{s}+\binom{K-1}{s-1}$, we have
\begin{align} 
\frac{\binom{K-2}{s-1}}{\binom{K-2}{s-1}+\sum_{i=0}^{K-1-s} \binom{K-1}{s+i}(N-1)^iN} &\geq 
\frac{\binom{K-2}{s-1}+\binom{K-2}{s-2}}{\binom{K-2}{s-1}+\binom{K-2}{s-2}+
\sum_{i=0}^{K-s}\left[ \binom{K-1}{s+i}+\binom{K-1}{s+i-1}   \right](N-1)^iN}, 
\end{align}
which is equivalent to
\begin{align} 
\frac{\binom{K-2}{s-1}}{\binom{K-2}{s-1}+\sum_{i=0}^{K-1-s} \binom{K-1}{s+i}(N-1)^iN} &\geq 
\frac{\binom{K-2}{s-2}}{\binom{K-2}{s-2}+\sum_{i=0}^{K-s}\binom{K-1}{s+i-1}(N-1)^iN}. \label{mm5}
\end{align}
By using \eqref{r_exp}, \eqref{mm5} is equivalent to 
\begin{align}
 r_{s}^{(K)} & \geq r_{s-1}^{(K)}.  \label{m2_4}
\end{align}
Since \eqref{m2_4} holds, we have $\alpha \geq 0$. Furthermore, $\alpha \leq 1$ can be proved similarly. For fixed $N$, since $\bar{D}(r^{(K+1)}_0) >  \bar{D}(r^{(K)}_0)$, the achievable normalized download cost monotonically increases. 
\end{Proof}

The following lemma provides an asymptotic upper bound for the achievable normalized download cost as a smooth function in $(r,N)$. From this expression, we characterize the worst-case gap between the outer and the inner bounds to be $\frac{1}{6}$.  
\begin{lemma}[Asymptotics and the worst-case gap]
	In cache-aided PIR with uncoded and unknown prefetching, as $K \rightarrow \infty$, the outer bound is tightly upper bounded by,
	\begin{align}
	\bar{D}(r) \leq \frac{N(1-r)^2}{(N-1)+r} 
	\end{align}
	Hence, the worst-case gap is $\frac{1}{6}$. The asymptotic unawareness multiplicative gain over memory-sharing in \cite{tandon2017capacity}  is $\frac{1-r}{1+\frac{r}{N-1}} \leq 1$.
\end{lemma}

\begin{Proof}
We write the outer bound $\bar{D}(r_s)$ as
\begin{align}  
\bar{D}(r_s)&= \frac{\sum_{i=0}^{K-1-s} \binom{K}{s+1+i}(N-1)^iN}{\binom{K-2}{s-1}+\sum_{i=0}^{K-1-s} \binom{K-1}{s+i}(N-1)^iN} \\
&=\frac{\frac{\sum_{i=0}^{K-1-s} \binom{K}{s+1+i}(N-1)^i}{\sum_{i=0}^{K-1-s} \binom{K-1}{s+i}(N-1)^i}}{\frac{\binom{K-2}{s-1}}{\sum_{i=0}^{K-1-s} \binom{K-1}{s+i}(N-1)^iN}+1}\\
&=\frac{\psi_1(N,K,s)}{\psi_2(N,K,s)+1}. \label{ub_gap}
\end{align}
Denote $\lambda=\frac{s}{K}$. To upper bound $\psi_1(N,K,s)$,
\begin{align}
\psi_1(N,K,s)&=\frac{\sum_{i=0}^{K-1-s} \binom{K}{s+1+i}(N-1)^i}{\sum_{i=0}^{K-1-s} \binom{K-1}{s+i}(N-1)^i} \\
&=\frac{\sum_{i=0}^{K-1-s} \frac{K}{s+1+i}\binom{K-1}{s+i}(N-1)^i}{\sum_{i=0}^{K-1-s} 
	\binom{K-1}{s+i}(N-1)^i}\\
& \leq \frac{\sum_{i=0}^{K-1-s} \frac{K}{s}\binom{K-1}{s+i}(N-1)^i}{\sum_{i=0}^{K-1-s} 
	\binom{K-1}{s+i}(N-1)^i}=\frac{1}{\lambda}.
\end{align}
	
We upper bound the reciprocal of $\psi_2(N,K,s)$ as,
\begin{align}
\frac{1}{\psi_2(N,K,s)}&=N\sum_{i=0}^{K-1-s} \frac{\binom{K-1}{s+i}(N-1)^i}{\binom{K-2}{s-1}}\\
&\label{gap_equality}=N\sum_{i=0}^{K-1-s} \frac{(K-1)(K-1-s)(K-2-s) \cdots (K-i-s)}{s(s+1)(s+2)\cdots (s+i)} (N-1)^i \\
&\leq N\sum_{i=0}^{K-1-s} \frac{K(K-s)^{i}}{s^{i+1}}(N-1)^i\\
&= N\sum_{i=0}^{(1-\lambda)K-1} \frac{(1-\lambda)^{i}}{\lambda^{i+1}}(N-1)^i\\
&=\frac{N}{\lambda} \sum_{i=0}^{(1-\lambda)K-1} \left(\frac{(1-\lambda)(N-1)}{\lambda}\right)^i. 
\end{align}
Now, if $\lambda > 1-\frac{1}{N}$, then $\frac{(1-\lambda)(N-1)}{\lambda} <1$. Hence, as $K \rightarrow \infty$, $\frac{1}{\psi_2(N,K,s)}$ converges to
\begin{align}
\lim_{K \rightarrow \infty} \frac{1}{\psi_2(N,K,s)} &\leq\frac{N}{\lambda} \sum_{i=0}^{\infty} \left(\frac{(1-\lambda)(N-1)}{\lambda}\right)^i \\ 
&=\frac{N}{\lambda}\cdot\frac{1}{1-\frac{(1-\lambda)(N-1)}{\lambda}} \\
&=\frac{N}{N\lambda-(N-1)}.
\end{align}
	
Moreover, \eqref{gap_equality} can be lower bounded by keeping the first $\epsilon K$ terms in the sum for any $\epsilon$ such that $0 < \epsilon <1-\lambda$,
\begin{align}
\frac{1}{\psi_2(N,K,s)} &\geq N\sum_{i=0}^{\epsilon K} \frac{(K-1)(K-1-s)(K-2-s) \cdots (K-i-s)}{s(s+1)(s+2)\cdots (s+i)}(N-1)^i \\
&\geq N\sum_{i=0}^{\epsilon K} \frac{(K-1)(K-\epsilon K-s)^{i}}{(s+\epsilon K)^{i+1}} (N-1)^i\\
&= N\sum_{i=0}^{\epsilon K} \frac{(1-\frac{1}{K})((1-(\lambda+\epsilon))^{i}}{(\lambda+\epsilon)^{i+1}}(N-1)^i.
\end{align}
Similarly, by taking $K \rightarrow \infty$, for any $0 < \epsilon <1-\lambda$, we have
\begin{align}
\lim_{K \rightarrow \infty} \frac{1}{\psi_2(N,K,s)} &\geq\frac{N}{\lambda+\epsilon} \sum_{i=0}^{\infty} \left(\frac{(1-(\lambda+\epsilon))(N-1)}{\lambda+\epsilon}\right)^i \\ 
&=\frac{N}{N(\lambda+\epsilon)-(N-1)}.
\end{align}
Since $\epsilon$ is arbitrarily chosen, then as $K \rightarrow \infty$, $\epsilon \rightarrow 0$, we have $\psi_2(N,K,s) \rightarrow \frac{N\lambda-(N-1)}{N}$.
	
Consequently, as $K \rightarrow \infty$, $r_s$ converges to
\begin{align} 
r_s \rightarrow r&= \lim_{K \rightarrow \infty}\frac{\binom{K-2}{s-1} }{\binom{K-2}{s-1}+\sum_{i=0}^{K-1-s} \binom{K-1}{s+i}(N-1)^i N} \\
&=\lim_{K \rightarrow \infty} \frac{\psi_2(N,K,s)}{\psi_2(N,K,s)+1}\\ 
&=\frac{N\lambda-(N-1)}{N\lambda+1}.
\end{align}
Note that if $\lambda=1-\frac{1}{N}$, then $r=0$, while if $\lambda=1$, then $r=\frac{1}{1+N}$. This means that the restriction in the limit to have $\lambda>1-\frac{1}{N}$ is without loss of generality as $\lambda>1-\frac{1}{N}$ corresponds to the entire range of $r$ other than the $1-r$ matching bound. We can write $\lambda$ as
\begin{align}
\lambda=\frac{r+(N-1)}{N(1-r)}. 
\end{align}
	
Substituting in \eqref{ub_gap}, we have the following upper bound on $\bar{D}(r)$
\begin{align}
\bar{D}(r) &\leq \frac{\frac{1}{\lambda}}{\frac{N\lambda-(N-1)}{N}+1} \\
&=\frac{N}{\lambda(N\lambda+1)} \\
&=\frac{N}{\frac{r+(N-1)}{N(1-r)}\left(\frac{r+(N-1)}{(1-r)}+1\right)}\\
&=\frac{N^2(1-r)^2}{(r+(N-1))^2+(1-r)(r+(N-1))} \\
&=\frac{N^2(1-r)^2}{Nr+N(N-1)}\\
&=\frac{N(1-r)^2}{(N-1)+r}. 
\end{align}
	
The memory-sharing scheme in \cite{tandon2017capacity} achieves $\frac{N}{N-1}(1-r)$ if $K \rightarrow \infty$, hence the asymptotic unawareness gain is given by the multiplicative factor $\frac{1-r}{1+\frac{r}{N-1}} \leq 1$. 
	
For the inner bound, we note that the $i$th corner point is given by,
\begin{align}
\tilde{r}_i=\frac{1}{1+N+\cdots+N^i}, \quad i=1,\cdots,K-1. 
\end{align}
Therefore, although there exist $K$ linear bounds, it suffices to consider only a small number of them, as the remaining bounds are concentrated around $r=0$. Denote the gap between the inner and the outer bounds by $\Delta(N,K,r)$.
We note that the gap $\Delta(N,\infty,r)$ is a piece-wise convex function for $0 \leq r \leq 1$ since it is the difference between a convex function $\bar{D}(r)$ and a piece-wise linear function. Hence, the maximizing caching ratio for the gap exists exactly at the corner points $\tilde{r}_i$ and it suffices to examine the gap at these corner points. 
	
For the outer bound, we have
\begin{align}
\bar{D}(\tilde{r}_i)&\leq \frac{N\left(1-\frac{1}{1+N+\cdots+N^i}\right)^2}{(N-1)+\frac{1}{1+N+\cdots+N^i}} \\
 &=\frac{N(1+N+N^2+\cdots+N^i-1)^2}{(N-1)(1+N+\cdots+N^i)^2+(1+N+\cdots+N^i)}\\
 &=\frac{N^2(1+N+\cdots+N^{i-1})^2}{N^{i}(1+N+\cdots+N^i)}. 
\end{align}
	
Furthermore, for the inner bound, we have
\begin{align}
\tilde{D}(\tilde{r}_i)&=\left(1+\frac{1}{N}+\cdots+\frac{1}{N^i}\right)-\frac{1}{1+N+\cdots+N^i}\left((i+1)+\frac{i}{N}+\cdots+\frac{1}{N^i}\right)\\
                  &=\frac{1+N+\cdots+N^i}{N^i}-\frac{(i+1)N^i+iN^{i-1}+\cdots+1}{N^i(1+N+\cdots+N^i)}\\
                  &=\frac{(1+N+\cdots+N^i)^2-(1+2N+3N^2+\cdots+(i+1)N^i)}{N^i(1+N+\cdots+N^i)}
\end{align}
	
	Consequently, we can upper bound the asymptotic gap at the corner point $\tilde{r}_i$ as
	\begin{align}
	\Delta(&N,\infty,\tilde{r}_i)    \notag \\
	&= \bar{D}(\tilde{r}_i)-\tilde{D}(\tilde{r}_i) \\
	                             &\leq \frac{N^2(1+N+\cdots+N^{i-1})^2-(1+N+\cdots+N^i)^2+(1+2N+3N^2+\cdots+(i+1)N^i)}{N^i(1+N+\cdots+N^i)}\\
	                             &=\frac{-(1+2N(1+N+\cdots+N^{i-1}))+(1+2N+3N^2+\cdots+(i+1)N^i)}{N^i(1+N+\cdots+N^i)} \\
	                             &=\frac{N^2+2N^3+\cdots+(i-1)N^i}{N^i(1+N+\cdots+N^i)}\\
	                             &=\frac{\frac{1}{N^{i-2}}+\frac{2}{N^{i-3}}+\cdots+(i-1)}{1+N+\cdots+N^i}
	\end{align}
    Hence, $\Delta(N,\infty,\tilde{r}_i)$ is monotonically decreasing in $N$. Therefore,
    \begin{align}
    \Delta(N,K,r) \leq \Delta(2,\infty,r) \leq \: \max_i \:\: \frac{(2)^2+2(2)^3+\cdots+(i-1)(2)^i}{2^i(1+2+\cdots+2^i)}
    \end{align}
    For the case $N=2$, we note that all the inner bounds after the $6$th corner point are concentrated around $r=0$ since $\tilde{r}_i \leq \frac{1}{127}$ for $i \geq 6$. Therefore, it suffices to characterize the gap only for the first $6$ corner points. Considering the $6$th corner point which corresponds to $\tilde{r}_6=\frac{1}{127}=0.0078$, and $\bar{D}(r) \leq 2$ trivially for all $r$, and $\tilde{D}(\frac{1}{127})=1.8898$. Hence, $\Delta(2,\infty,r) \leq 0.11$, for $r \leq \frac{1}{127}$. Now, we focus on calculating the gap at $\tilde{r}_i$, $i=1, \cdots, 6$. Examining all the corner points, we see that $r=\frac{1}{15}$ is the maximizing caching ratio for the gap (corresponding to $i=3$), and $\Delta(2,\infty,\frac{1}{15}) \leq \frac{1}{6}$, which is the worst-case gap.  
\end{Proof}

\section{Conclusion}
In this paper, we studied the cache-aided PIR problem from $N$ non-communicating and replicated databases, when the cache stores uncoded bits that are unknown to the databases. We determined inner and outer bounds for the optimal normalized download cost $D^*(r)$ as a function of the total number of messages $K$, the number of databases $N$, and the caching ratio $r$. Both inner and outer bounds are piece-wise linear functions in $r$ (for fixed $N$, $K$) that consist of $K$ line segments. The bounds match in two specific regimes: the very low caching ratio regime, i.e., $r \leq \frac{1}{1+N+N^2+\cdots+N^{K-1}}$, where $ D^*(r)=(1-r)\left(1+\frac{1}{N}+\cdots+\frac{1}{N^{K-1}}\right)-r\left((K-1)+\frac{K-2}{N}+\cdots+\frac{1}{N^{K-2}}\right)$; and the very high caching ratio regime, where $D^*(r)=(1-r)(1+\frac{1}{N})-r$, for $\frac{K-2}{(N+1)K+N^2-2N-2}\leq r\leq \frac{1}{1+N}$ and $D^*(r)=1-r$, for $r\geq \frac{1}{1+N}$. As a direct corollary for this result, we characterized the exact tradeoff between the download cost and the caching ratio for $K=3$. For general $K$, $N$, and $r$, we showed that the largest gap between the achievability and the converse bounds is $\frac{1}{6}$. The outer bound shows significant reduction in the download cost with respect to the case when the cache content is fully known at all databases \cite{tandon2017capacity}, which achieves $D^*(r)=(1-r)(1+\frac{1}{N}+\cdots+\frac{1}{N^{K-1}})$ by memory-sharing. 

The achievable scheme extends the greedy scheme in \cite{JafarPIR} so that it starts with exploiting the cache bits as side information. For fixed $K$, $N$, there are $K-1$ non-degenerate corner points. These points differ in the number of cached bits that contribute in generating one side information equation. The achievability for the remaining caching ratios is done by memory-sharing between the two adjacent corner points that enclose that caching ratio $r$. For the converse, we extend the induction-based techniques in \cite{JafarPIR,tandon2017capacity} to account for the availability of uncoded and unknown prefetching at the retriever. The converse proof hinges on developing $K-1$ lower bounds on the length of the undesired portion of the answer string. By applying induction on each bound separately, we obtain the piece-wise linear inner bound.

\bibliographystyle{unsrt}
%\bibliography{reference}
\bibliography{references}
\end{document}